\pdfoutput=1

\documentclass[11pt,twoside,a4paper,cmspaper,final,collab]{cms-tdr}
\def\svnVersion{487184P}\def\cmsCernNoTag{CERN-EP-2018-167}\def\cmsCernDate{\today}\def\cmsMessage{Published in the European Physical Journal C as \href{http://dx.doi.org/10.1140/epjc/s10052-018-6482-9}{\doi{10.1140/epjc/s10052-018-6482-9.}}}

\begin{document}\cmsNoteHeader{SMP-16-003}

\hyphenation{had-ron-i-za-tion}
\hyphenation{cal-or-i-me-ter}
\hyphenation{de-vices}
\RCS$HeadURL: svn+ssh://svn.cern.ch/reps/tdr2/papers/SMP-16-003/trunk/SMP-16-003.tex $
\RCS$Id: SMP-16-003.tex 487097 2019-01-24 02:03:23Z rslu $
\newlength\cmsFigWidth
\ifthenelse{\boolean{cms@external}}{\setlength\cmsFigWidth{0.98\columnwidth}}{\setlength\cmsFigWidth{0.60\textwidth}}
\newlength\cmsFigWidthBig
\ifthenelse{\boolean{cms@external}}{\setlength\cmsFigWidthBig{0.55\textwidth}}{\setlength\cmsFigWidthBig{0.65\textwidth}}
\ifthenelse{\boolean{cms@external}}{\providecommand{\cmsLeft}{top\xspace}}{\providecommand{\cmsLeft}{left\xspace}}
\ifthenelse{\boolean{cms@external}}{\providecommand{\cmsRight}{bottom\xspace}}{\providecommand{\cmsRight}{right\xspace}}
\newcommand{\specialcell}[2][c]{
  \begin{tabular}[#1]{@{}c@{}}#2\end{tabular}}
\newcommand{\JETPHOX}{\textsc{jetphox}\xspace}
\newlength\cmsTabSkip\setlength{\cmsTabSkip}{1ex}
\ifthenelse{\boolean{cms@external}}{\providecommand{\cmsTable}[1]{#1}}{\providecommand{\cmsTable}[1]{\resizebox{\textwidth}{!}{#1}}}
\cmsNoteHeader{SMP-16-003}
\title{Measurement of differential cross sections for inclusive isolated-photon and photon+jet production in proton-proton collisions at \texorpdfstring{$\sqrt{s} = 13\TeV$}{sqrt(s) = 13 TeV}}
\titlerunning{Differential cross sections for inclusive isolated-photon and photon+jet production at 13\TeV}

\date{\today}

\abstract{Measurements of inclusive isolated-photon and photon+jet production in proton-proton collisions at $\sqrt{s} = 13\TeV$ are presented. The analysis uses data collected by the CMS experiment in 2015, corresponding to an integrated luminosity of 2.26\fbinv.
The cross section for inclusive isolated photon production is measured as a function of the photon transverse energy in a fiducial region.  The cross section for photon+jet production is measured as a function of the photon transverse energy in the same fiducial region with identical photon requirements and with the highest transverse momentum jet. All measurements are in agreement with predictions from next-to-leading-order perturbative QCD.}

\hypersetup{
pdfauthor={CMS Collaboration},
pdftitle={Measurement of differential cross sections for inclusive isolated-photon and photon+jet production in proton-proton collisions at sqrt(s) = 13 TeV},
pdfsubject={CMS},
pdfkeywords={CMS, physics, SMP}}

\maketitle

\section{Introduction}
The measurement of inclusive isolated-photon and photon+jet production cross sections can directly probe quantum chromodynamics (QCD).
The dominant production processes in proton-proton (\Pp\Pp) collisions at the energies of the CERN LHC
are quark-gluon Compton scattering $\cPq\cPg \to \cPq\Pgg$, together with contributions from quark-antiquark annihilation $\cPq\Paq \to \cPg\Pgg$,
and parton fragmentation $\cPq\cPaq (\cPg\cPg) \to X+\Pgg$.
Both the CMS and ATLAS Collaborations have reported measurements of the differential cross sections for isolated prompt photon
production~\cite{Khachatryan:2010fm,Chatrchyan:2011ue,Chatrchyan:2012vq,Aad:2010sp,Aad:2011tw,Aad:2013zba, Aad:2016xcr}
and for the production of a photon in association with jets~\cite{Chatrchyan:2013mwa, ATLAS:2012ar, Aaboud:2016sdm}
using data with center-of-mass energies of 2.76, 7, and 8\TeV.
The ATLAS Collaboration has also reported the same measurements at a center-of-mass energy of 13\TeV~\cite{Aaboud:2017cbm, Aaboud:2017kff}.

The published measurements show agreement with the results of next-to-leading-order (NLO) perturbative QCD calculations~\cite{Aurenche:2006vj,Ichou:2010wc}.

These LHC measurements are sensitive to the gluon density function $\cPg(x, Q^2)$
over a wide range of parton momentum fraction $x$ and energy scale $Q^2$~\cite{Vogelsang:1995bg, d'Enterria:2012yj, Carminati:2012mm}.
These measurements were not included in the global parton distribution function (PDF)
fits~\cite{Ball:2014uwa,Harland-Lang:2014zoa,Dulat:2015mca} until very recently~\cite{Campbell:2018wfu}.
An improved understanding of all PDFs is key to reducing the associated theoretical uncertainties in
the calculation of many relevant cross sections, including Higgs boson production and new physics searches.

In this paper, measurements are reported for the inclusive isolated-photon cross section in a fiducial region using data collected by the CMS Collaboration in proton-proton collisions at $\sqrt{s} = 13\TeV$, corresponding to an integrated luminosity of 2.26\fbinv~\cite{CMS:2016eto}. The specific fiducial region is defined at generator level as: (1) photon transverse momentum $\et > 190\GeV$, (2) rapidity $\abs{y} < 2.5$, and (3) an isolated photon where the sum of the \pt of all particles inside a cone of radius $\Delta R = \sqrt{\smash[b]{(\Delta\phi)^2 + (\Delta\eta)^2}} = 0.4$ around the photon is less than 5\GeV. The photon+jet cross section is also measured in this fiducial region with the same photon requirements and with $\pt^{\text{jet}}>30\gev$ and $\abs{{y}^{\text{jet}}}<2.4$. The significant increase in center-of-mass energy compared with the previous CMS papers~\cite{Khachatryan:2010fm,Chatrchyan:2011ue} opens a large additional region of phase space.

The dominant background for the photon+jet process is QCD multijet production with an isolated electromagnetic (EM) deposit from decays of neutral hadrons, mostly from \Pgpz~mesons.
A multivariate analysis method is used to identify prompt photons using a boosted decision tree (BDT) algorithm, implemented using the TMVA v4.1.2 toolkit~\cite{Hocker:2007ht}. Photon yields are extracted using the shape of the BDT distributions,
and the measured cross sections are compared to the results of NLO QCD calculations.

\section{The CMS detector}
CMS is a general-purpose detector built to explore physics at the TeV scale.
The central feature of the CMS apparatus is a superconducting solenoid of 6\unit{m} internal diameter,
providing a magnetic field of 3.8\unit{T}. Within the solenoid volume are a silicon pixel and a strip tracker,
a lead tungstate crystal electromagnetic calorimeter (ECAL), and a brass and scintillator hadron calorimeter (HCAL),
each composed of a barrel and two endcap sections. Forward calorimeters extend the pseudorapidity $\eta$ coverage provided by the barrel and endcap detectors.
Muons are measured in gas-ionization detectors embedded in the steel flux return yoke outside the solenoid.
A more detailed description of the CMS detector, together with the definition of the coordinate system and the relevant kinematic variables,
is given in Ref.~\cite{Chatrchyan:2008aa}.

The ECAL consists of 75\,848 lead tungstate crystals, which provide coverage up to $\abs{ \eta } = 1.479 $ in the barrel region (EB) and $1.479 < \abs{ \eta } < 3.0$ in two endcap regions (EE). A preshower detector consisting of two planes of silicon sensors interleaved with a total of $3$ radiation lengths of lead is located in front of the EE.

The silicon tracker measures charged particles within the range $\abs{\eta} < 2.5$.
For nonisolated particles of transverse momenta $1 < \pt < 10\GeV$ and $\abs{\eta} < 1.4$, the track resolutions are typically 1.5\% in \pt and 25--90 (45--150)\mum in the transverse (longitudinal) impact parameter~\cite{Chatrchyan:2014fea}.

 The global event reconstruction (also called particle-flow event reconstruction)~\cite{CMS-PRF-14-001} reconstructs and identifies each particle candidate
with an optimized combination of all subdetector information.

In CMS, both converted and unconverted photons are reconstructed using ECAL clusters and are included in the analysis.
The clustering algorithm results in an almost complete collection of the energy of the photons, unconverted ones and those converting in the material upstream
of the calorimeter. First, cluster ``seeds" are identified as local energy maxima above a given threshold. Second,
clusters are grown from the seeds by aggregating crystals with at least one side in common with a clustered crystal and with an energy in excess of a given threshold.
This threshold represents about two standard deviations of the electronic noise, which depends on $\abs{\eta}$.
The energy in an individual crystal can be shared between clusters under the assumption that each seed corresponds to a single EM particle.
Finally, clusters are merged into ``superclusters", to allow good energy containment, accounting for geometrical variations of the
detector along $\eta$, and increasing robustness against additional \Pp\Pp\xspace collisions in the same or adjacent bunch crossings (pileup).
The clustering excludes $1.44 < \abs{ \eta } <1.56$, which corresponds to the transition region between the EB and EE.
The fiducial region terminates at $\abs{\eta}=2.5$ where the tracker coverage ends.

The energy of photons is computed from the sum of the energies of the clustered crystals, calibrated and corrected for degradation in the crystal response over time~\cite{Chatrchyan:2013dga}. The preshower energy is added to that of the superclusters in the region covered by this detector.
To optimize the resolution, the photon energy is corrected using a multivariate regression technique that estimates the containment of the electromagnetic shower in the superclusters, the shower losses for photons that convert in the material upstream of the calorimeter, and the effects of pileup~\cite{CMS:EGM-14-001}.
The regression training is performed on simulated events using shower shape and position variables of the photon as inputs.
The regression provides a per-photon estimate of the function parameters that quantify the containment, the shower losses, and pileup and therefore a prediction of the distribution of the ratio of true energy to the uncorrected supercluster energy. The most probable value of this distribution is taken as the photon energy correction.
The regression output is used to correct the reconstucted photon energy in data to agree with simulated events. An additional smearing is applied to the photon energy in simulation to reproduce the resolution observed in data. The scale correction and smearing procedure uses a multistep procedure exploiting electrons from $\cPZ \to \EE$ decays.
In the EB, an energy resolution of about 1\% is achieved for unconverted photons in the tens of \GeV energy range. The remaining EB photons have a resolution of about 1.3\% up to $\abs{\eta}=1.0$, rising to about 2.5\% at $\abs{\eta}=1.4$. In the EE, the resolution of unconverted or late-converting photons is about 2.5\%, while the remaining EE photons have a resolution between 3 and 4\%.

Electrons are identified as a primary charged track consistent with potentially multiple ECAL energy clusters from both the electron and from potential bremsstrahlung photons produced in the tracker material.
Muons are identified as a track in the central tracker consistent with either a track or several hits in the muon system, associated with a minimum ionization signature in the calorimeters.
Charged hadrons are charged-particle tracks not identified as electrons or muons.
Finally, neutral hadrons are identified as HCAL energy clusters not linked to any charged-hadron track, or as ECAL and HCAL energy excesses with respect to the expected charged-hadron energy deposit.

Jets are clustered from all particle candidates reconstructed by the global event reconstruction with the infrared- and collinear- safe anti-\kt
algorithm~\cite{Cacciari:2008gp, Cacciari:2011ma} using a distance parameter $R$ of 0.4.
The momenta of jets reconstructed using particle-flow candidates in the simulation are within 5 to 10\% of particle-level jet momenta over the whole jet \pt spectrum and detector acceptance, and corrected on average accordingly. In situ measurements of the momentum balance in dijet, photon+jet, $\cPZ$+jet, and multijet events are used to correct for any residual differences in jet energy scale in data and simulation~\cite{Khachatryan:2016kdb}. The jet energy resolution amounts typically to 15 (8)\% at 10 (100)\GeV.

\section{Simulation samples}
Simulated event samples for photon+jet and multijet final states are generated at leading order (LO) with \PYTHIA~8 (v8.212)~\cite{Sjostrand:2014zea}.
The photon+jet sample contains direct photon production originating from quark-gluon Compton scattering and quark-antiquark annihilation.

The multijet sample, which is dominated by final states with quark and gluon jets, is used in the estimate of systematic uncertainties, and to estimate the small bias in the extracted photon yield from the BDT fit, as described in section~\ref{sec:CS}. For these studies, events containing a photon, produced via the fragmentation process and passing the fiducial requirements, are removed, leaving only events with nonfiducial photons. The removed events are considered part of the signal, although they are not included in the signal sample in the training of the BDT due to associated large statistical uncertainties.  The distributions of the variables used in the BDT training were examined and are consistent with those of the direct photons, within the statistical uncertainty.

{\tolerance=1000
The \MADGRAPH~(v5.2.2.2)~\cite{Alwall:2014hca, Alwall:2007fs} LO generator, interfaced with \PYTHIA~8, is used to generate an additional sample of photon+jet
events containing up to 4 jets that are used to estimate systematic uncertainties.
Samples of $\cPZ/\gamma^*$+jets events are generated at NLO with \MGvATNLO~(v5.2.2.2)~\cite{Alwall:2014hca, Frederix:2012ps} and are used for calibration and validation studies described later.
The CUETP8M1 tune~\cite{Khachatryan:2015pea} is used in \PYTHIA~8.
The NNPDF2.3 LO PDF~\cite{Ball:2012cx} and the NNPDF3.0 NLO PDF~\cite{Ball:2014uwa} are used to generate simulation samples, where the former is used with \PYTHIA~8.
\par}

The simulated processes include the effect of the pileup.
The pileup contribution is simulated with additional minimum bias events superimposed on the primary event
using the measured distribution of the number of reconstructed interaction vertices, an average of 14 vertices per bunch crossing.
A detailed detector simulation based on the \GEANTfour~(v9.4p03)~\cite{Agostinelli:2002hh} package is applied to all the generated signal and background samples.

\section{Data samples and event selection criteria}
Events containing high energy photon candidates are selected using the two-level CMS trigger system~\cite{Khachatryan:2016bia}. At the first level, events are accepted if they have an ECAL trigger tower, which has a segmentation corresponding to $5\times 5$ ECAL crystals, with total transverse energy \et, defined as the magnitude of the photon transverse momentum, greater than 40\GeV.
The second level of the trigger system uses the same reconstruction algorithm as the offline photon reconstruction~\cite{CMS:EGM-14-001}.
An event is accepted online if it contains at least one ECAL cluster with \et greater than 175\GeV, and if the ``$H/E$", defined as the ratio of energy deposited in the HCAL to that in the ECAL, is less than 0.15 (0.10) in the EB (EE) region.

{\tolerance=1000
All events are required to have at least one well-reconstructed primary vertex~\cite{Chatrchyan:2014fea}.
The reconstructed vertex with the largest value of summed physics-object $\pt^2$ is the primary $\Pp\Pp$ interaction vertex.
The physics objects are the jets, clustered using the jet finding algorithm~\cite{Cacciari:2008gp, Cacciari:2011ma} with the tracks assigned to the vertex as inputs, and the associated missing transverse momentum $\ptmiss$~\cite{CMS:2016ljj}, taken as the negative vector sum of the \pt of those jets.
In addition, photon+jet events are required to be balanced in \pt, and hence the magnitude of missing transverse momentum, defined as the magnitude of the negative vector sum of the momenta of all reconstructed particle-flow objects projected onto the plane perpendicular to the beam axis in an event,
is required to be less than 70\% of the highest photon \et.
\par}

Photon candidates are selected as described in the following procedure.
An electron veto is imposed by requiring the absence of hits in the innermost layer of the silicon pixel detector
that could be ascribed to an electron track consistent with the energy and position of the photon ECAL cluster.
Criteria on the energy measured in HCAL ($H$), isolation, and shower shape variables are applied to reject photons arising from electromagnetic decays of particles in hadronic showers.
Hence, $H/E$ is required to be less than 0.08 (0.05) for photon candidates in the EB (EE), respectively.
The sum of the \et of other photons in a cone (photon isolation) of size $\Delta R = 0.3$ around the photon candidate
is required to be less than 15\GeV, and the sum of \pt of charged hadrons in the same cone (hadron isolation)
is required to be less than 2.0 (1.5)\GeV for photon candidates in the EB (EE).

To further suppress photons from decays of neutral mesons ($\pi^0$, $\eta$, \etc)
that survive the isolation and HCAL energy leakage criteria, a selection on the EM shower shape is imposed by requiring that its second moment
$\sigma_{\eta \eta}$~\cite{CMS:EGM-14-001}, which is a measure of the lateral extension of the shower along the $\eta$ direction,
be $<$0.015\,(0.045) for photon candidates in the EB\,(EE).
The photon candidate with the highest \et that satisfies the above selection criteria in each event is referred to as the leading photon.
The data consist of 212\,134 events after applying inclusive isolated-photon selections and 207\,120 events after applying the photon+jet requirements. The estimated electron contribution is typically at $10^{-3}$ level as a result of the electron veto algorithm. This contribution is small compared to statistical uncertainties of the photon yield and other systematic uncertainties.

The photon reconstruction and selection efficiencies are estimated using simulated events that pass the fiducial region requirements at the generator level.
The efficiency is about 90--92\% (83--85\%) for EB (EE) photons, depending on the \et of the photon candidate.
The loss of efficiency comes primarily from the hadron isolation requirement.
Multiplicative scale factors (SF) are applied to correct potential differences in efficiencies between data and simulation.
The SFs are obtained from the ratio of the efficiency in data to that in simulated control samples.
The photon SF is derived from Drell--Yan $\cPZ \to \EE$ events, where one of the electrons is reconstructed as a photon.
The events are selected by requiring the invariant mass of the electron pair to be between 60--120\GeV.
The electron veto SF is determined using final-state radiation photons in $\cPZ\to \MM\gamma$ events.
All SFs are within 1\% of unity, and their uncertainties are included in the total systematic uncertainty.
All efficiencies and SF are measured as functions of photon \et and rapidity $y$ using the same binning as the cross section measurement.

The absolute photon trigger efficiency, as a function of photon \et,
is measured using events collected with a jet trigger that contains a photon candidate, which satisfies the signal selection criteria and is spatially separated from the jet that triggered the event by $\Delta R (\gamma, \text{jet}) > 0.7$.
The trigger efficiency is above 99\% for EB\,(EE) photons above 200\,(220)\GeV. The \et-dependent trigger efficiency is used to compute the cross section,
and the associated uncertainties are incorporated into the uncertainty calculation for the cross section.

For the cross section measurement as a function of jet $y$, the jets are required to: (1) satisfy a set of
selection criteria that remove detector noise~\cite{CMS:2017wyc}, (2) have a separation from
the leading photon of $\Delta R > 0.4$, and (3) have \pt greater than 30\GeV.
The \pt requirement for jets is fully efficient for simulation events with both photon and jet in their fiducial regions.
The jet candidate with the highest \pt satisfying the above requirements is selected.

The measurement of the differential cross section for inclusive isolated photons uses four ranges of photon rapidity,
$\abs{{y}^{\gamma}}<0.8$, $0.8<\abs{{y}^{\gamma}}<1.44$, $1.57<\abs{{y}^{\gamma}}<2.1$, and $2.1<\abs{{y}^{\gamma}}<2.5$.
The photon+jet differential cross section measurement uses two ranges of photon rapidity,
$\abs{{y}^{\gamma}}<1.44$ and $1.57<\abs{{y}^{\gamma}}<2.5$,
and two ranges of jet rapidity, $\abs{{y}^{\text{jet}}}<1.5$ and $1.5<\abs{{y}^{\text{jet}}}<2.4$.
For all cases, the results are presented in nine bins in photon \et between 190 to 1000\GeV, except for two cases:
the $2.1<\abs{{y}^{\gamma}}<2.5$ region for the isolated-photon measurement and the $1.57<\abs{{y}^{\gamma}}<2.5$ and
$1.5<\abs{{y}^{\text{jet}}}<2.4$ regions for the photon+jet measurement, where eight bins in photon \et between 190 to 750\GeV are used.

\section{Cross section measurement}\label{sec:CS}
To further suppress remaining backgrounds originating from jets faking photons, a BDT is constructed utilizing the following discriminating variables:
\begin{enumerate}
\item Photon $\eta$, $\phi$, and energy;
\item Several shower shape variables:
\begin{enumerate}
\item The energy sum of the $3\times 3$ crystals centered on the most energetic crystal in the photon divided by the energy of the photon;
\item The ratio of $E_{2\times 2}$, the maximum energy sum collected in a $2\times 2$ crystal matrix that includes the largest energy crystal in the photon, and $E_{5\times 5}$, the energy collected in a $5\times 5$ crystal matrix centered around the same crystal ($E_{2\times 2}/E_{5\times 5}$);
\item The second moment of the EM cluster shape along the $\eta$ direction ($\sigma_{\eta\eta}$);
\item The diagonal component of the covariance matrix that is constructed from the energy-weighted crystal positions within the $5\times 5$ crystal array ($q_{\eta \phi}$);
\item The energy-weighted spreads along $\eta$ ($\sigma_{\eta}$) and $\phi$ ($\sigma_{\phi}$), calculated using all crystals in the photon cluster, which provide further measures of the lateral spread of the shower.
\end{enumerate}
\item For photon candidates in the EE, the preshower shower width, $\sigma_{RR} = \sqrt{\smash[b]{\sigma_{xx}^2 + \sigma_{yy}^2}}$, where $\sigma_{xx}$ and $\sigma_{yy}$ measure the lateral spread in the two orthogonal sensor planes of the detector, and the fraction of energy deposits in the preshower.
\item The median energy density per unit area in the event $\rho$~\cite{Cacciari:2011ma} to minimize the effect of the pileup.
\end{enumerate}

The distributions of the BDT values are used in a two-template binned likelihood fit to estimate the photon yield.
A separate BDT is constructed for each bin of photon $y$ and \et. The signal BDT template is obtained from the sample of simulated photon+jet events generated using \PYTHIA~8.
This template is validated using $\cPZ\to\MM \gamma$ data samples and also a data sample of $\cPZ\to \EE$ candidates where each candidate contains an electron reconstructed as a photon.
The signal templates have a systematic uncertainty due to differences in the distributions of the BDT input variables in data and simulation.
To evaluate this uncertainty, the distribution of each variable obtained from a sample of simulated $\cPZ\to \EE$ events is modified until agreement is obtained with the data.
Signal templates are made using the same procedure.
The difference in the templates is treated as a nuisance parameter in the fit procedure.

\begin{figure}[th]
\centering
\includegraphics[width=\cmsFigWidth]{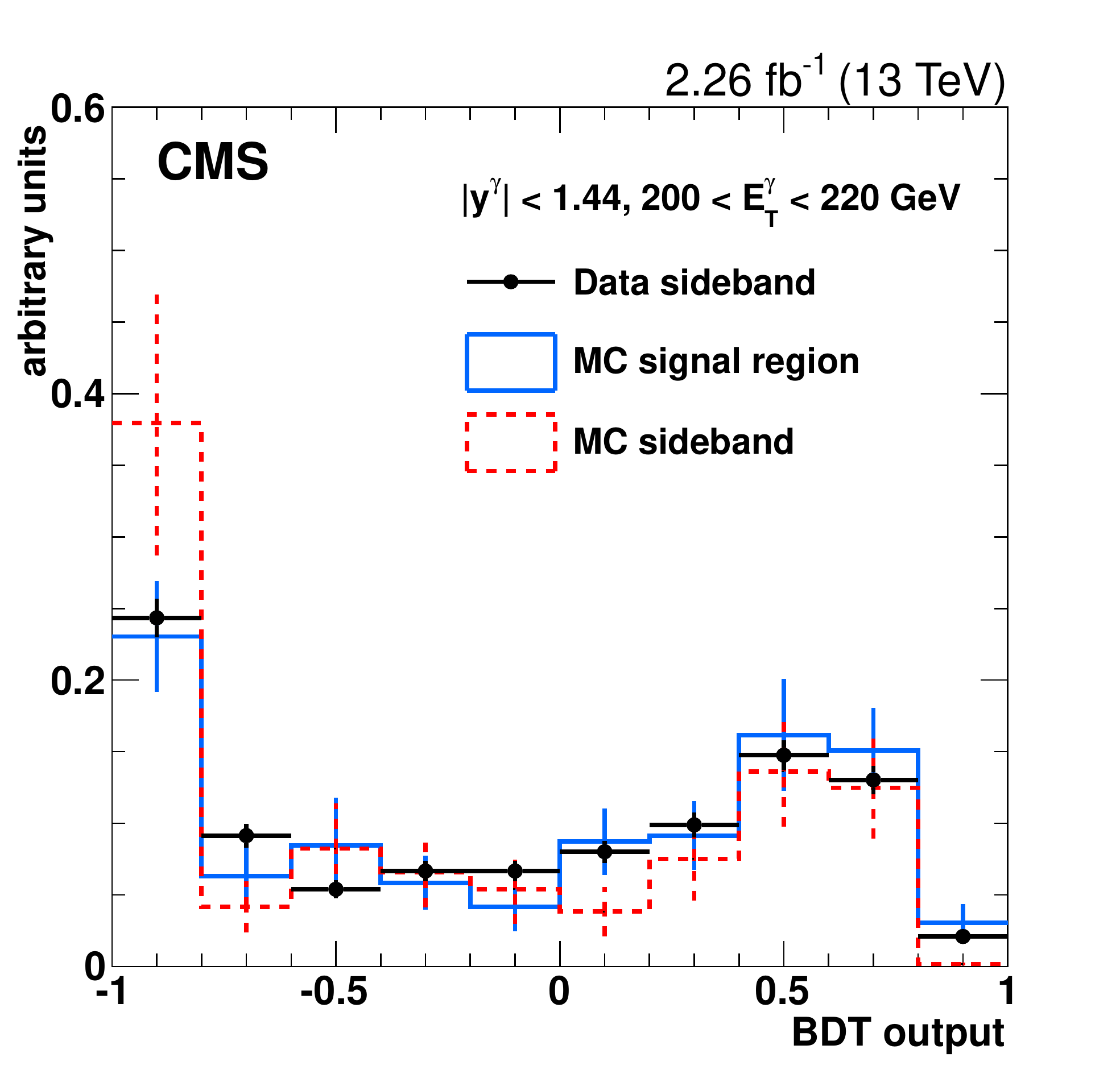}
\caption{Distributions of the BDT for background photons in the 200--220\GeV bin for the EB region.
The points show events from a sideband region of the photon isolation selection criteria,
the solid histogram shows the events in the signal region in simulated QCD multijet events,
and the dashed histogram shows the sideband region for simulated QCD multijet events.
All three samples have their statistical uncertainties shown as error bars.}
\label{fig:bkg_template}
\end{figure}

\begin{figure*}[htb!]
\includegraphics[width=0.48\textwidth]{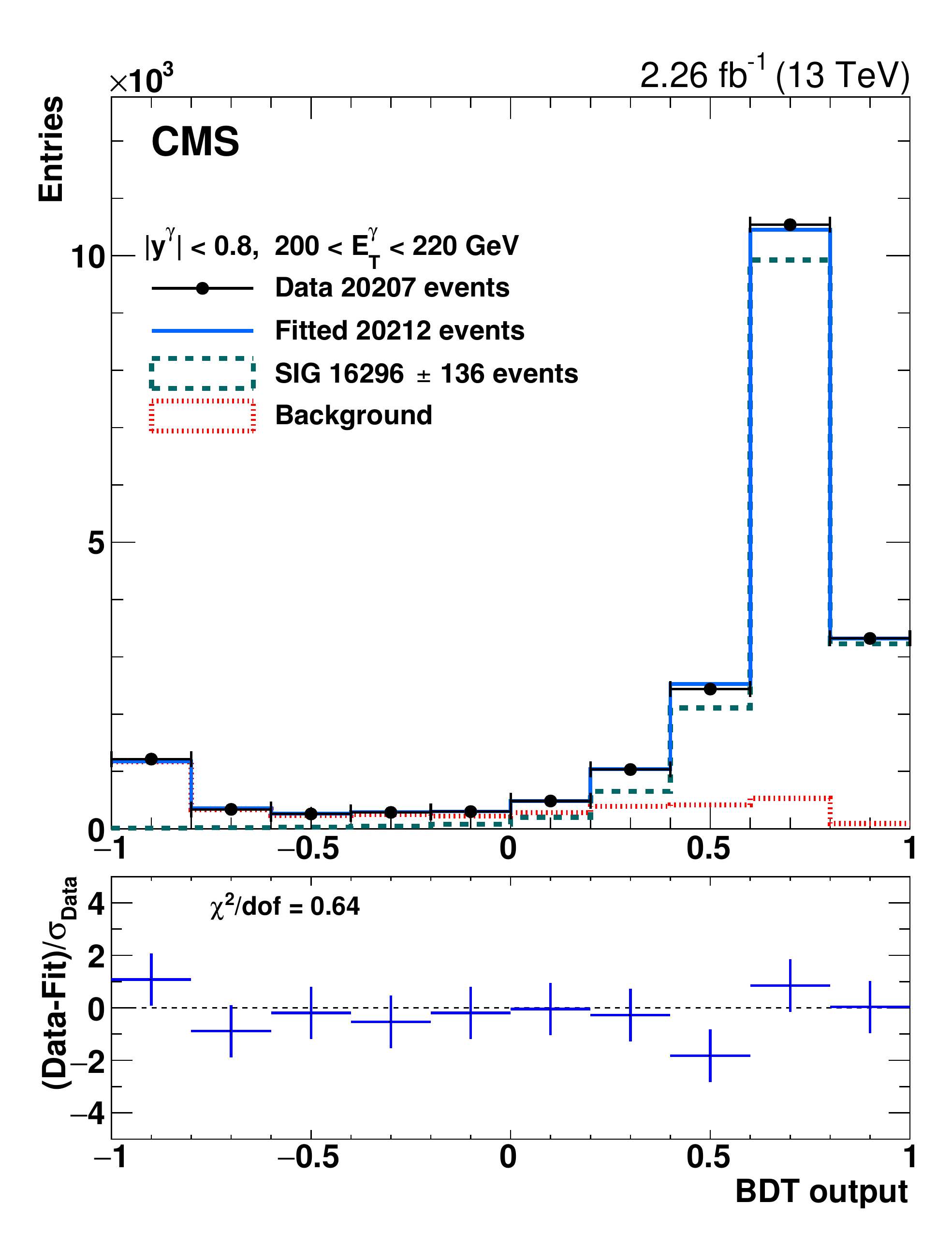}
\includegraphics[width=0.48\textwidth]{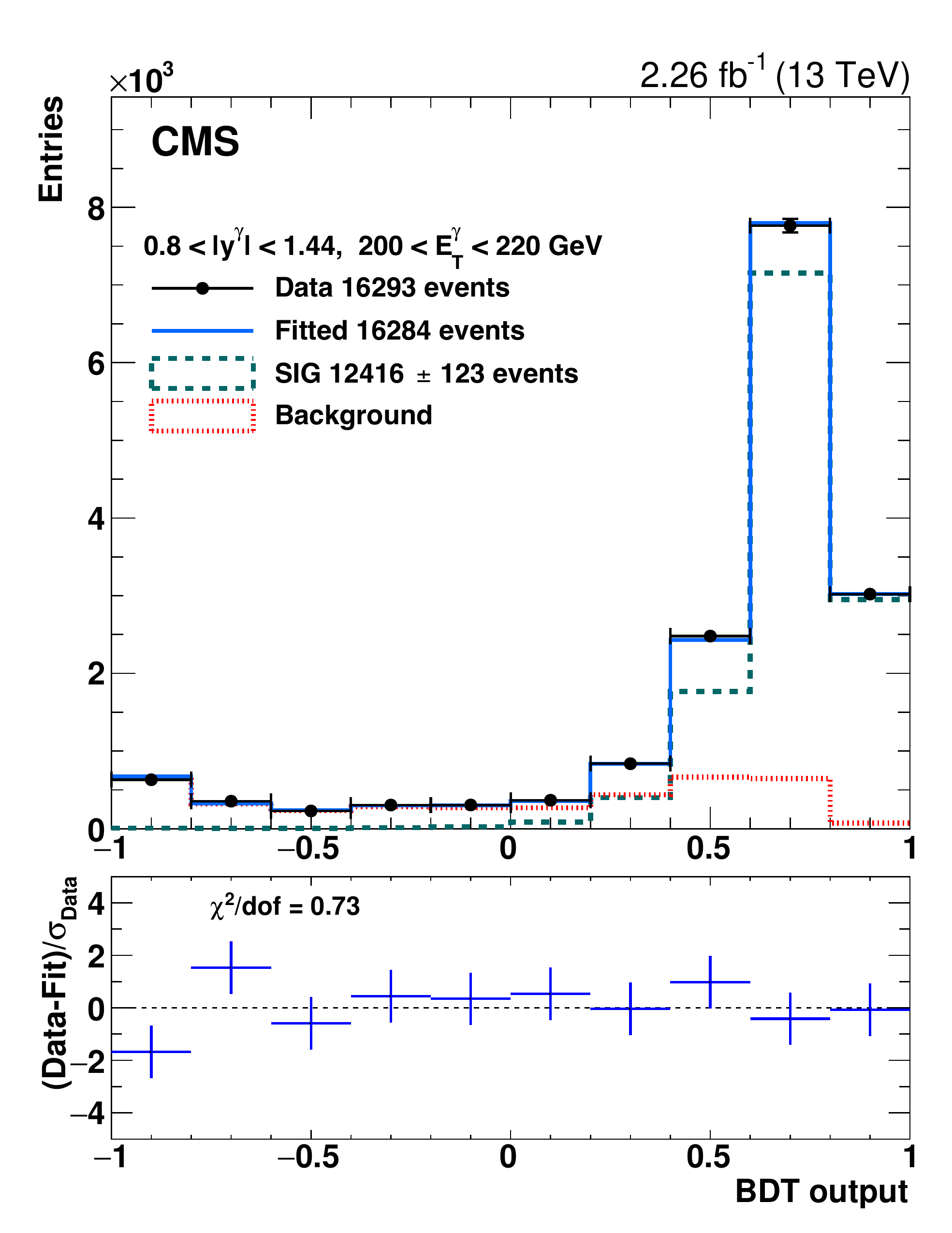}
\caption{Distributions of the BDT output for an EB (left) and an EE (right) bin with photon \et between 200--220\GeV and $\abs{ y^{\text{jet}} }<1.5$.
The points represent data, and the solid histograms, approaching the data points,
represent the fit results with the signal (dashed) and background (dotted) components displayed.
The bottom panels show the ratio of the data to the fitted results and the $\chi^2$/dof.}
\label{fig:mL_fit}
\end{figure*}

The background BDT template is derived from the data, using a sideband region defined using the same signal selection, but relaxing the hadron isolation criterion.
The hadron isolation for the sideband region is required to be between 7 and 13 (6 and 12)\GeV for EB (EE) photons, where the chosen ranges ensure negligible signal contamination.
Possible biases in the photon yields due to differences between the background BDT templates in the control and signal regions are estimated using simulated events and are found to be less than 5\%.
Photon yields extracted from the fits are corrected for these biases.
The statistical uncertainties in each bin of the background template constructed from the data sideband events are also included as nuisance parameters in the fitting procedure.
Figure~\ref{fig:bkg_template} shows the BDT templates obtained for a particular photon \et and $y$ bin for the data sideband and for the signal and sideband regions from simulated QCD multijet events.
The distributions of BDT outputs for EB and EE photons in data are shown in Fig.~\ref{fig:mL_fit} for photon \et between 200 and 220\GeV and jet $\abs{y} <1.5$.
The fitted results for the signal, background, and combined distributions are also shown in Fig.~\ref{fig:mL_fit}.  The ratio of experimental data to the simulation results demonstrates agreement as indicated by the $\chi^2$ per degree of freedom.

The corrected signal yield is unfolded using the iterative D'Agostini
method~\cite{DAgostini:1994fjx}, as implemented in the RooUnfold software package~\cite{Adye:2011gm}, to take into account migrations between different bins due to the photon energy scale and resolution, and into and out of the fiducial \et region. The unfolding response matrix is obtained from the \PYTHIA~8 photon+jet sample. The unfolding corrections are small, of the order of 1\%.
The size of the corrections is also verified using an independent photon+jet sample generated with \MADGRAPH.

The inclusive isolated-photon differential production cross section is calculated as
\begin{linenomath*}
\begin{equation}
\frac{\rd^2\sigma}{\rd y^{\gamma} \rd \et^{\gamma}} = \frac{{\mathcal U}(N^{\gamma})}{\Delta y^{\gamma}\Delta\et^{\gamma} }\;\frac{1}{\epsilon \, \text{SF} \, L},
\end{equation}
\end{linenomath*}
and the photon+jet as
\begin{linenomath*}
\begin{equation}
\frac{\rd^3\sigma}{\rd y^{\gamma} \rd \et^{\gamma} \rd {y}^{\text{jet}}} = \frac{{\mathcal U}(N^{\gamma})}{\Delta y^{\gamma}\Delta\et^{\gamma}\Delta {y}^{\text{jet}} }\;\frac{1}{\epsilon \, \text{SF} \, L},
\end{equation}
\end{linenomath*}
where $\mathcal U(N^{\gamma})$ denotes the unfolded photon yields in bins of width $\Delta \et^{\gamma}$ and $\Delta y$,
and $y$ is the rapidity of either the photon or the jet.
In these equations, $\epsilon$ denotes the product of trigger, reconstruction, and selection efficiencies;
$\text{SF}$ the product of the selection and electron veto scale factors; and $L$ is the integrated luminosity.

\section{Systematic uncertainties}
The uncertainty in the efficiency of the event selection is typically small except in the high-\et region,
where statistical uncertainties in both data and simulated events dominate.
A summary of the systematic uncertainties in the cross section measurement, due to the uncertain in  trigger and event selection efficiencies, Data-to-MC scale factors, signal and background template shapes, bin migrations from the unfolding procedure, and uncertainties in the photon energy scale and resolution, is given in Table~\ref{tab:systematics}. All of the above are treated as uncorrelated.

The systematic uncertainties in the trigger efficiency are dominated by the statistical uncertainty in jet trigger data where the trigger efficiencies are measured. The uncertainties of the selection efficiency are dominated by the statistical uncertainties of the simulation sample. The uncertainties of the Data-to-MC scale factor are based on the available $\cPZ \to \EE$ events, and a \pt extrapolation is employed.

The systematic uncertainties in the signal and background templates are incorporated into the fit as nuisance parameters.
For the signal template uncertainty, the nuisance parameter is assigned a Gaussian prior,
while log-normal priors are assigned to the background template nuisances.
A description of the general methodology can be found in Ref.~\cite{LHC-HCG}.
The bias correction, applied to the photon yields, due to the selection of the sideband range is also considered as a systematic uncertainty.

The impact on photon yields due to the event migration between photon \pt bins from the unfolding uncertainties, which include photon energy scale and resolution uncertainties, is roughly 5\%.
The uncertainties of the event selection efficiency due to the jet selection, jet energy scale and resolution, and jet rapidity migration are negligible.

The total uncertainty, not considering luminosity uncertainty, in the yield per bin, excluding the highest photon \et bin in each $y$ range, is about 5--8\% for EB and 9--17\% for EE photons.
The highest photon \et bins in all $y$ region have limited events in data and simulated samples for the evaluation of systematics.

The uncertainty in the measurement of the CMS integrated luminosity is 2.3\%~\cite{CMS:2016eto} and it is added in quadrature with other systematic uncertainties.

\begin{table*}[h]
\centering
\topcaption{Impact on cross sections, in percent, for each systematic uncertainty source in the four photon rapidity regions,
$\abs{y^{\gamma}} < 0.8$,  $0.8< \abs{y^{\gamma}} < 1.44$,  $1.57 < \abs{y^{\gamma}} < 2.1$, and $2.1 < \abs{y^{\gamma}} < 2.5$.
The ranges, when quoted, indicate the variation over photon \et between 190--1000\GeV.}\label{tab:systematics}
\begin{tabular}{lcccc} \hline
Source & $\abs{y^{\gamma}} < 0.8$ & $0.8< \abs{y^{\gamma}} < 1.44$ & $1.57 < \abs{y^{\gamma}} < 2.1$ & $2.1 < \abs{y^{\gamma}} < 2.5$  \\ \hline
Trigger efficiency & 0.7--8.5 & 0.2--13.4 & 0.6--20.5 & 0.3--7.8 \\
Selection efficiency & 0.1--1.3 & 0.1--1.3 & 0.1--5.3 & 0.1--1.1 \\
Data-to-MC scale factor & 3.7 & 3.7 & 7.1 & 7.1\\
Template shape & 0.6--5.0 & 0.1--10.2 & 0.5--4.9 & 0.6--16.2\\
Event migration & 3.8--5.5 & 1.2--4.1 & 2.0--8.5 & 2.3--10.3\\
Total w/o luminosity& 5.4--12.0 & 5.9--18.2 & 8.2--26.9 & 8.6--21.7\\
Integrated luminosity & \multicolumn{4}{c}{2.3} \\\hline
\end{tabular}
\end{table*}

\section{Results and comparison with theory}
\begin{figure*}[htb!]
\centering
\includegraphics[width=\cmsFigWidthBig]{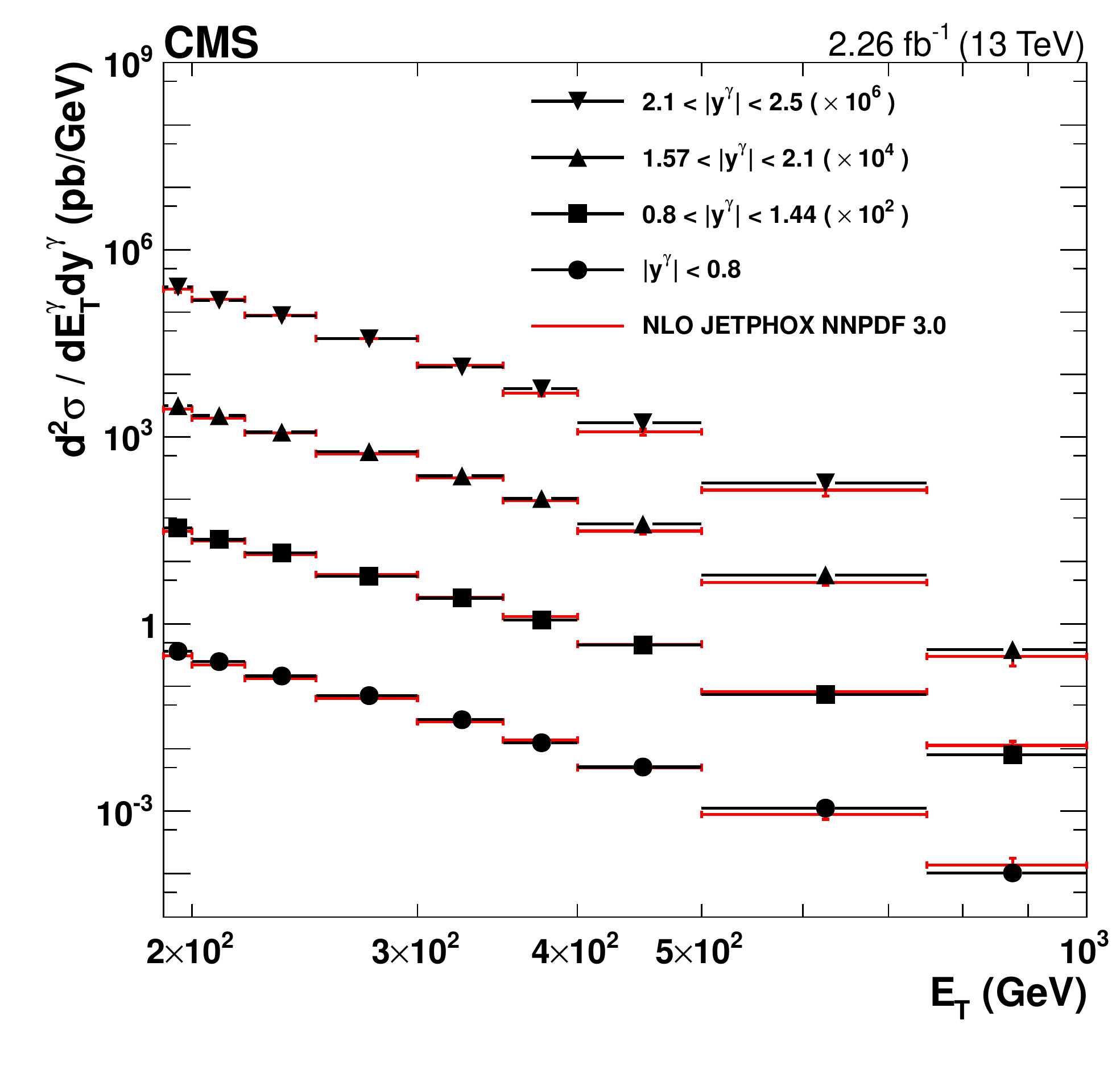}
\caption{Differential cross sections for isolated-photon production in photon rapidity bins, $\abs{y^{\gamma}} < 0.8$,  $0.8< \abs{y^{\gamma}} < 1.44$,  $1.57 < \abs{y^{\gamma}} < 2.1$, and $2.1 < \abs{y^{\gamma}} < 2.5$.
The points show the measured values and their total uncertainties; the lines show the NLO \JETPHOX~predictions with the NNPDF3.0 PDF set.}
\label{fig:EBEE_XS}
\end{figure*}

\begin{figure*}[htb!]
\centering
\includegraphics[width=0.48\textwidth]{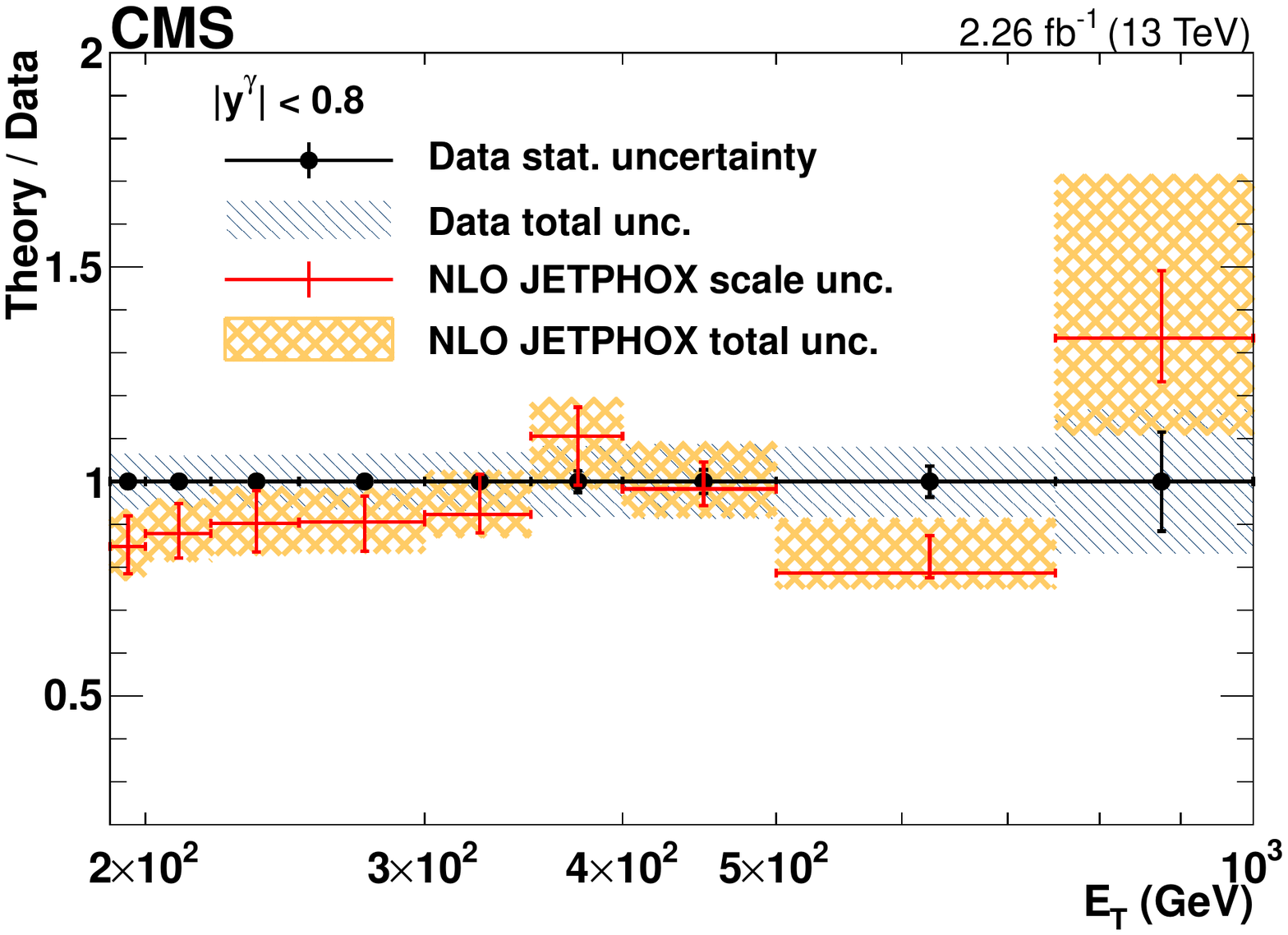}
\includegraphics[width=0.48\textwidth]{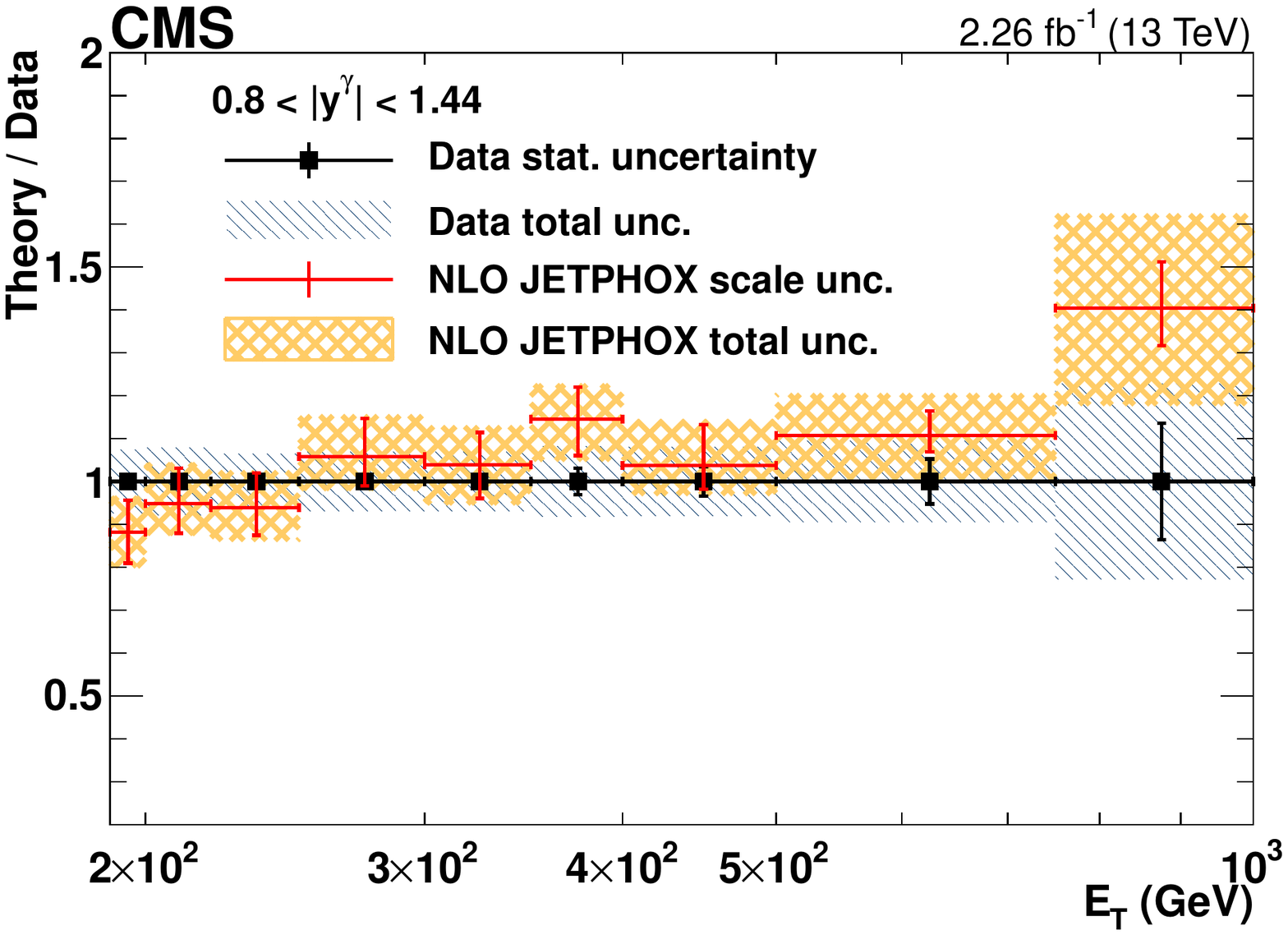}
\includegraphics[width=0.48\textwidth]{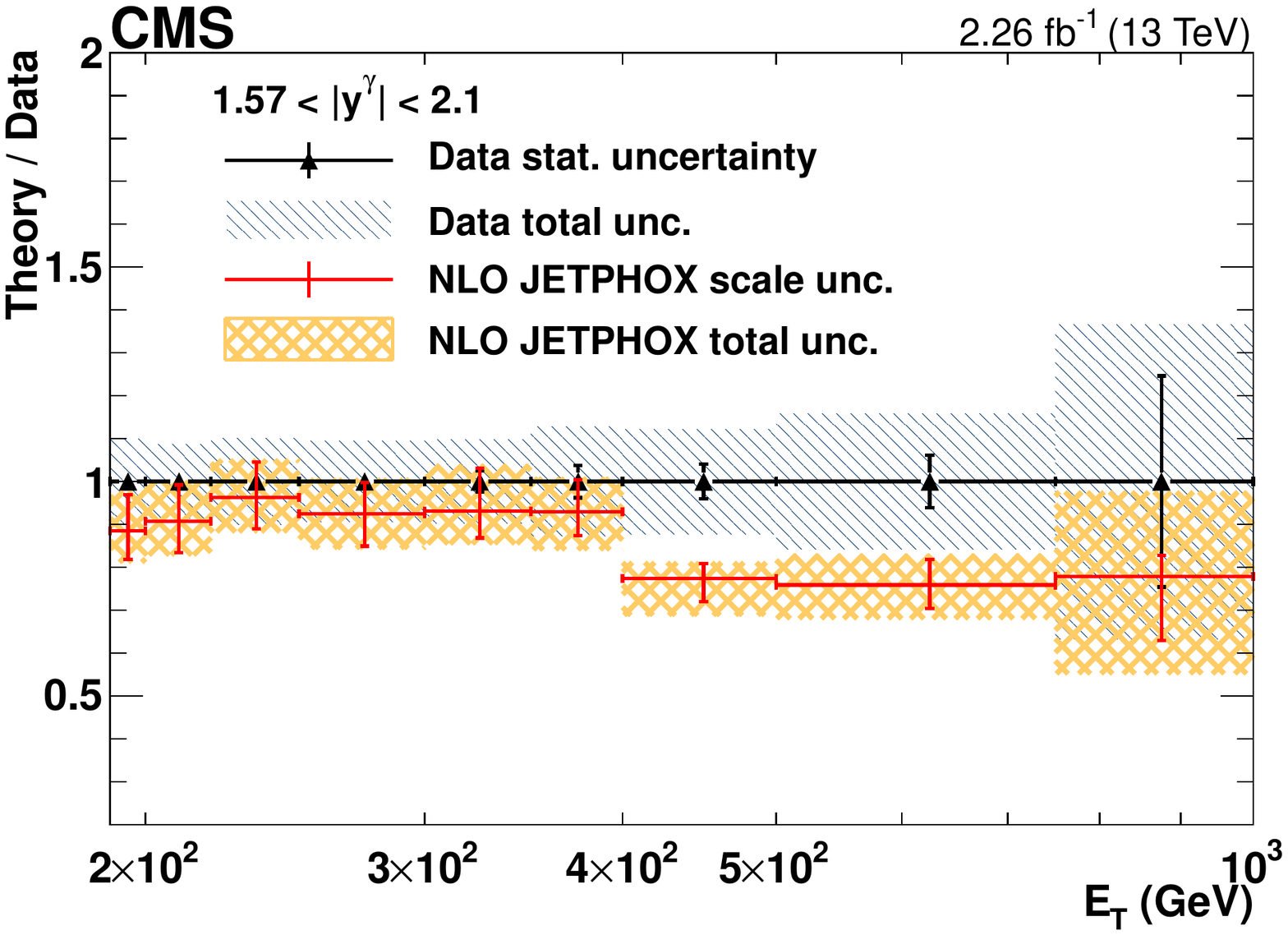}
\includegraphics[width=0.48\textwidth]{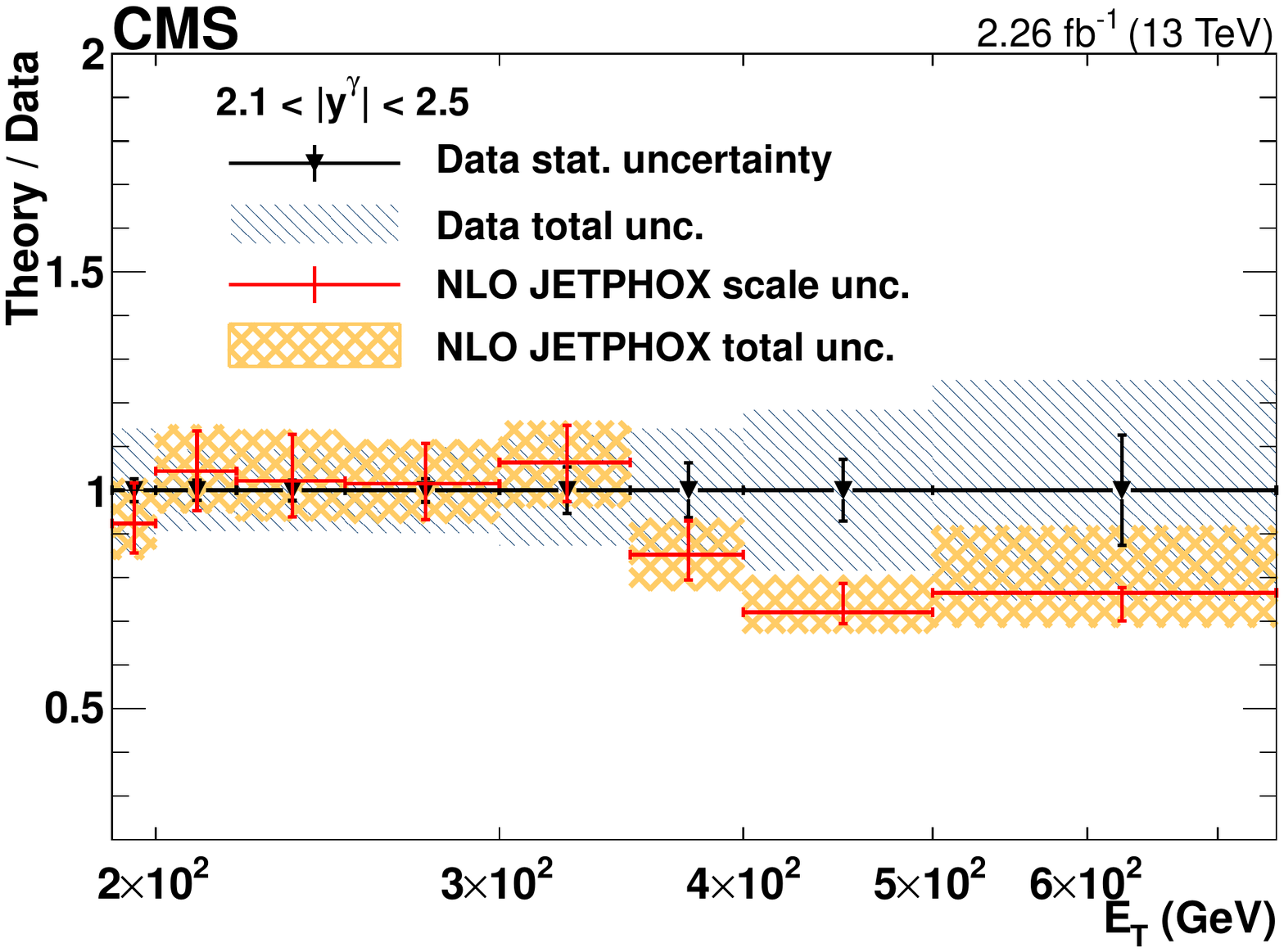}
\caption{The ratios of theoretical NLO predictions to data for the differential cross sections for isolated-photon production in four photon rapidity bins,
$\abs{y^{\gamma}} < 0.8$,  $0.8< \abs{y^{\gamma}} < 1.44$,  $1.57 < \abs{y^{\gamma}} < 2.1$, and $2.1 < \abs{y^{\gamma}} < 2.5$, are shown.
The error bars on data points represent the statistical uncertainty, while the hatched area shows the total experimental uncertainty.
The errors on the ratio represent scale uncertainties, and the shaded regions represent the total theoretical uncertainties.}
\label{fig:EBEE_XS_ratio}
\end{figure*}

\begin{figure*}[htb!]
\centering
\includegraphics[width=\cmsFigWidthBig]{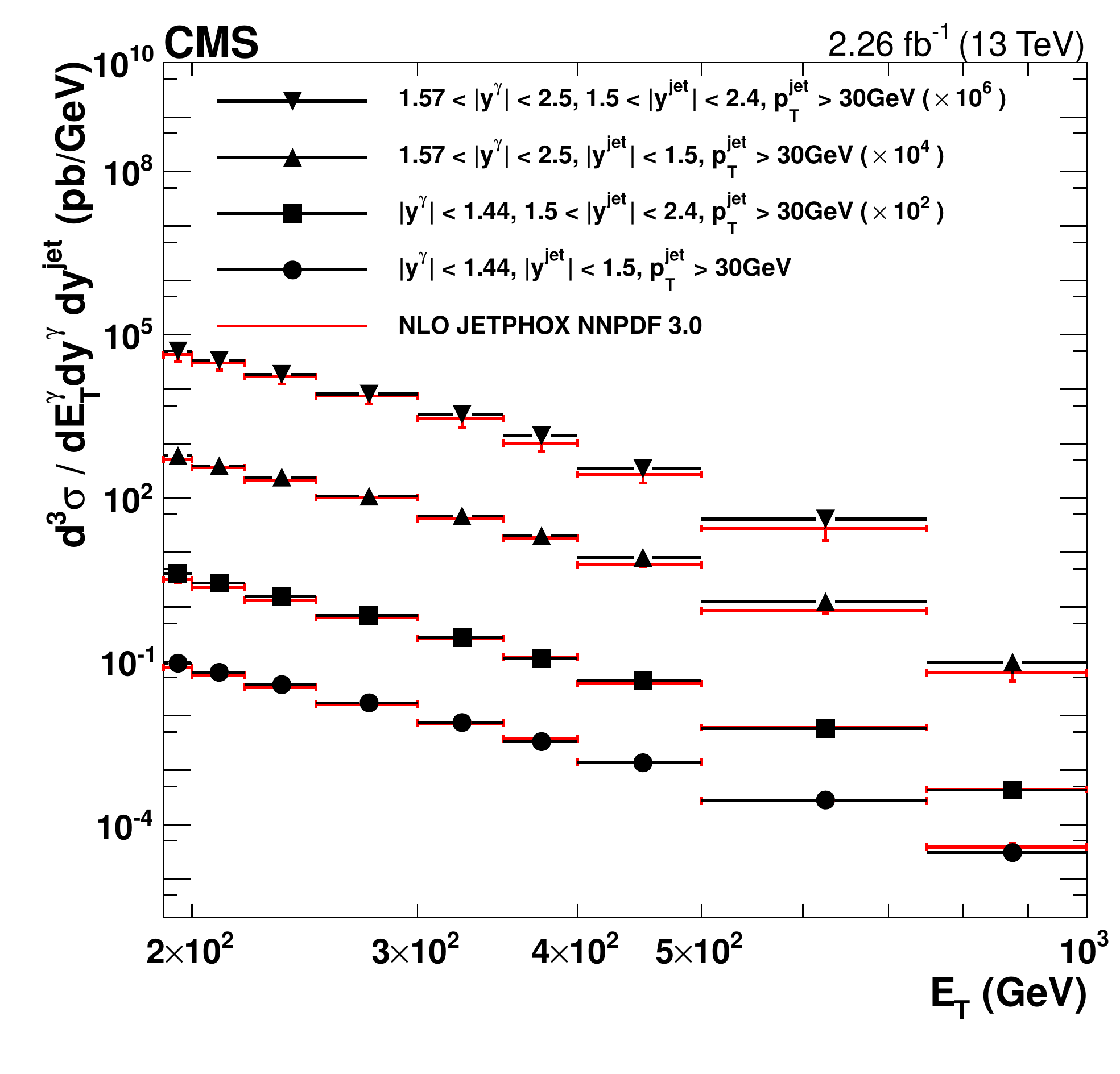}
\caption{Differential cross sections for photon+jet production in two photon rapidity bins, $\abs{y^{\gamma}} < 1.44$ and $1.57 < \abs{y^{\gamma}} < 2.5$, and two jet rapidity bins, $\abs{y^{\text{jet}}} < 1.5$ and $1.5 < \abs{y^{\text{jet}}} < 2.4$.
The points show the measured values with their total uncertainties, and the lines show the NLO \JETPHOX~predictions with the NNPDF3.0 PDF set.}
\label{fig:phojet_XS}
\end{figure*}

\begin{figure*}[htb!]
\centering
\includegraphics[width=0.48\textwidth]{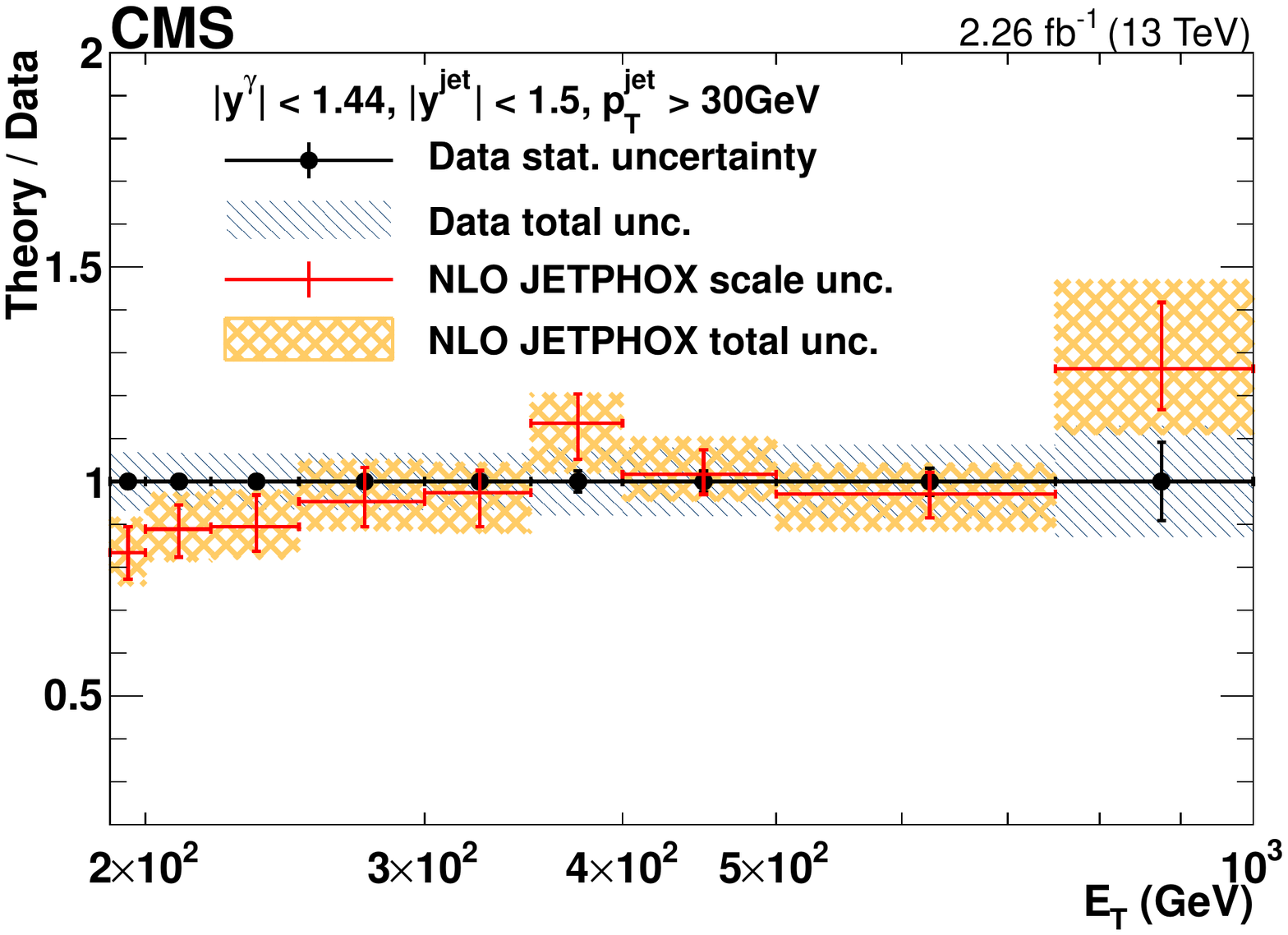}
\includegraphics[width=0.48\textwidth]{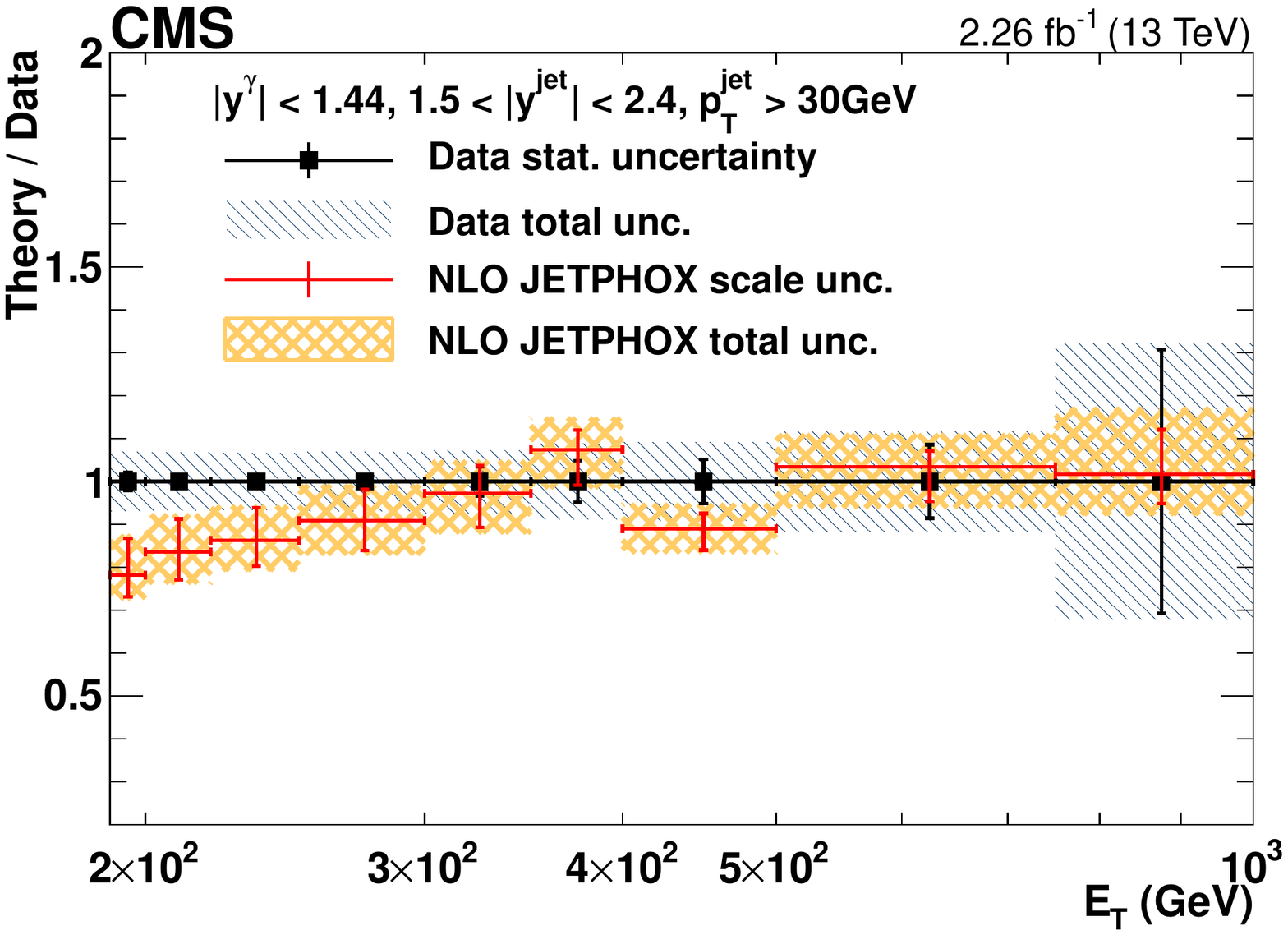}
\includegraphics[width=0.48\textwidth]{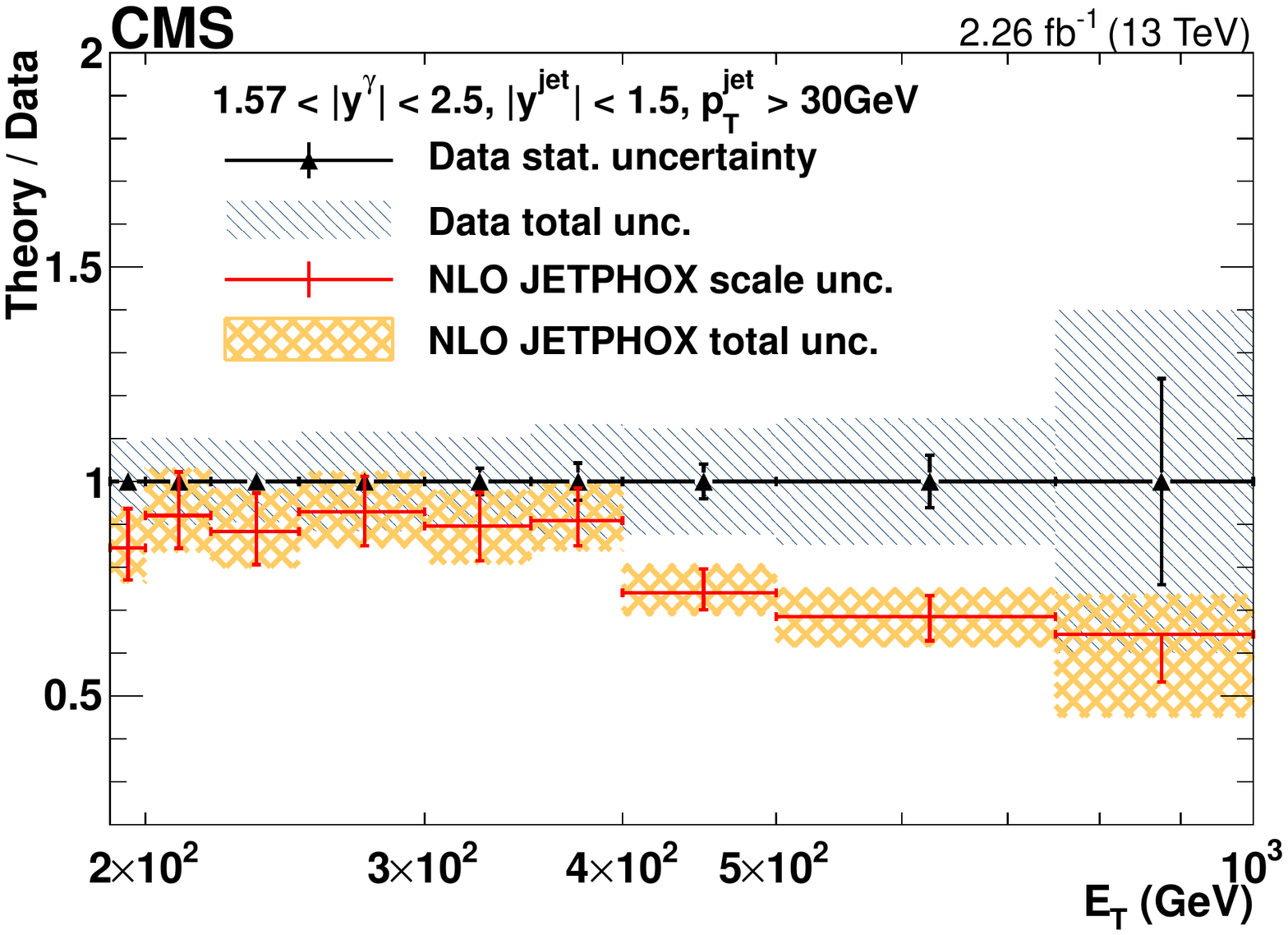}
\includegraphics[width=0.48\textwidth]{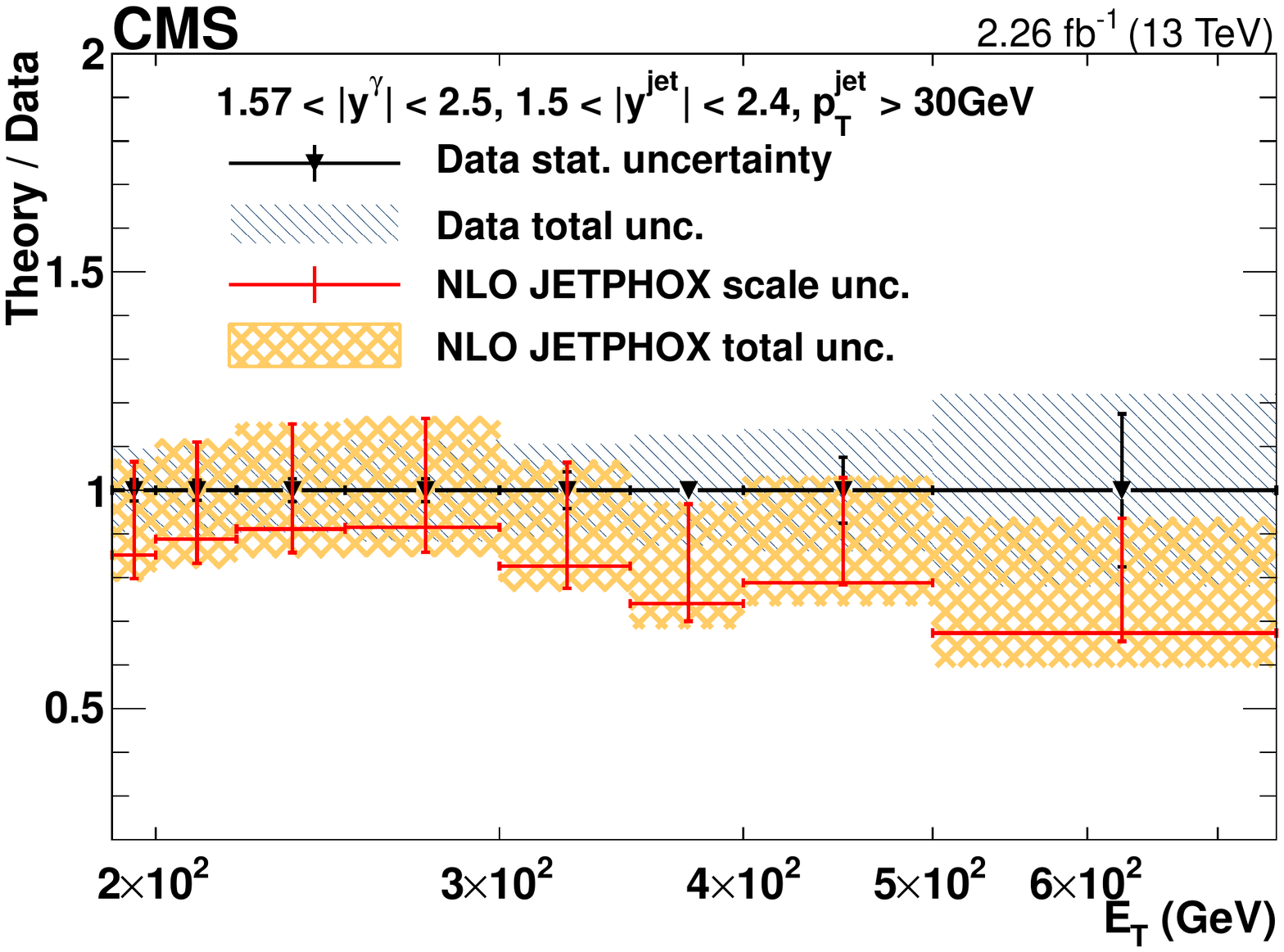}
\caption{The ratios of theoretical NLO prediction to data for
the differential cross sections for photon+jet production in two photon rapidity ($\abs{y^{\gamma}} < 1.44$ and $1.57 < \abs{y^{\gamma}} < 2.5$)
and two jet rapidity ($\abs{y^{\text{jet}}} < 1.5$ and $1.5 < \abs{y^{\text{jet}}} < 2.4$) bins , are shown.
The error bars on the data points represent their statistical uncertainty,  while the hatched area shows the total experimental uncertainty.
The error bars on the ratios show the scale uncertainties, and the shaded area shows the total theoretical uncertainties.}
\label{fig:phojet_ratio}
\end{figure*}

The measured inclusive isolated-photon cross sections as a function of photon \et are shown in Fig.~\ref{fig:EBEE_XS}
and the ratio compared with theory in Fig.~\ref{fig:EBEE_XS_ratio}
for photon \et greater than 190\gev and $\abs{y^{\gamma}} < 2.5 $ in 4 rapidity bins. The results are listed in Table~\ref{tab:summary_xs}.
The measurements for photon+jet cross sections as a function of photon \et are shown in Fig.~\ref{fig:phojet_XS} and the ratio compared with theory in Fig.\ref{fig:phojet_ratio} with additional requirements of $\pt^{\text{jet}}>30\gev$ and $|y^{\text{jet}}|<2.4$. The results are binned in two photon rapidity and two jet rapidity bins and are listed in Table~\ref{tab:summary_xs_phojet}.
The predictions require an isolated photon at generator level as described previously, with a transverse isolation energy less than 5\GeV.

The measured cross sections in the overlapping photon ET regions are increased by approximately a factor of 3 to 5 compared to previous CMS measurements at 7\TeV~\cite{Khachatryan:2010fm, Chatrchyan:2011ue, Chatrchyan:2013mwa}. This 13\TeV analysis also extends the photon \et range from 400 (300)\GeV in the 7 \TeV inclusive photon (photon+jet) results to 1\TeV.

The measured cross sections are compared with NLO perturbative QCD calculations from the \JETPHOX~1.3.1 generator~\cite{Catani:2002ny, Aurenche:2006vj, Belghobsi:2009hx}, using the NNPDF3.0 NLO~\cite{Ball:2014uwa} PDFs and the Bourhis-Fontannaz-Guillet (BFG) set II parton fragmentation functions~\cite{Bourhis:1997yu}.
The renormalization, factorization, and fragmentation scales are all set to be equal to the photon \et. To estimate the effect of the choice of theoretical scales on the predictions, the three scales are varied independently from $\et/2$ to $2\et$, while keeping their ratio between one-half and two.
The impact of \JETPHOX~cross section predictions due to the uncertainties in the PDF and in the
strong coupling $\alpS = 0.118$ at the mass of \PZ boson is calculated using the 68\% confidence level NNPDF3.0 NLO replica.
The uncertainty of parton-to-particle level transformation of the NLO pQCD prediction due to the underlying event and parton shower is studied by comparing with dedicated \PYTHIA samples where the choice and tuning of the generator has been modified.
The differences between the dedicated \PYTHIA and the nominal sample are between 0.5 to 2.0\%, depending on the photon \et and $y$, and they are assigned as the systematic uncertainty.
The total theoretical uncertainties of the cross section predictions are evaluated as the quadratic sum of the scale, PDF,\alpS, and underlying event and parton shower uncertainties.

The ratio of the theoretical predictions to data, together with the experimental and theoretical uncertainties, are shown in Figs.~\ref{fig:EBEE_XS_ratio} and~\ref{fig:phojet_ratio} for the isolated-photon and photon+jet cross section measurements respectively.
The uncertainties in the theoretical predictions and ratios to data are symmetrized in the tables; the largest value between the positive and negative uncertainties is listed.
Measured cross sections are in agreement with theoretical expectations within statistical and systematic uncertainties.

\begin{figure*}[htbp]
\centering
\includegraphics[width=0.42\textwidth]{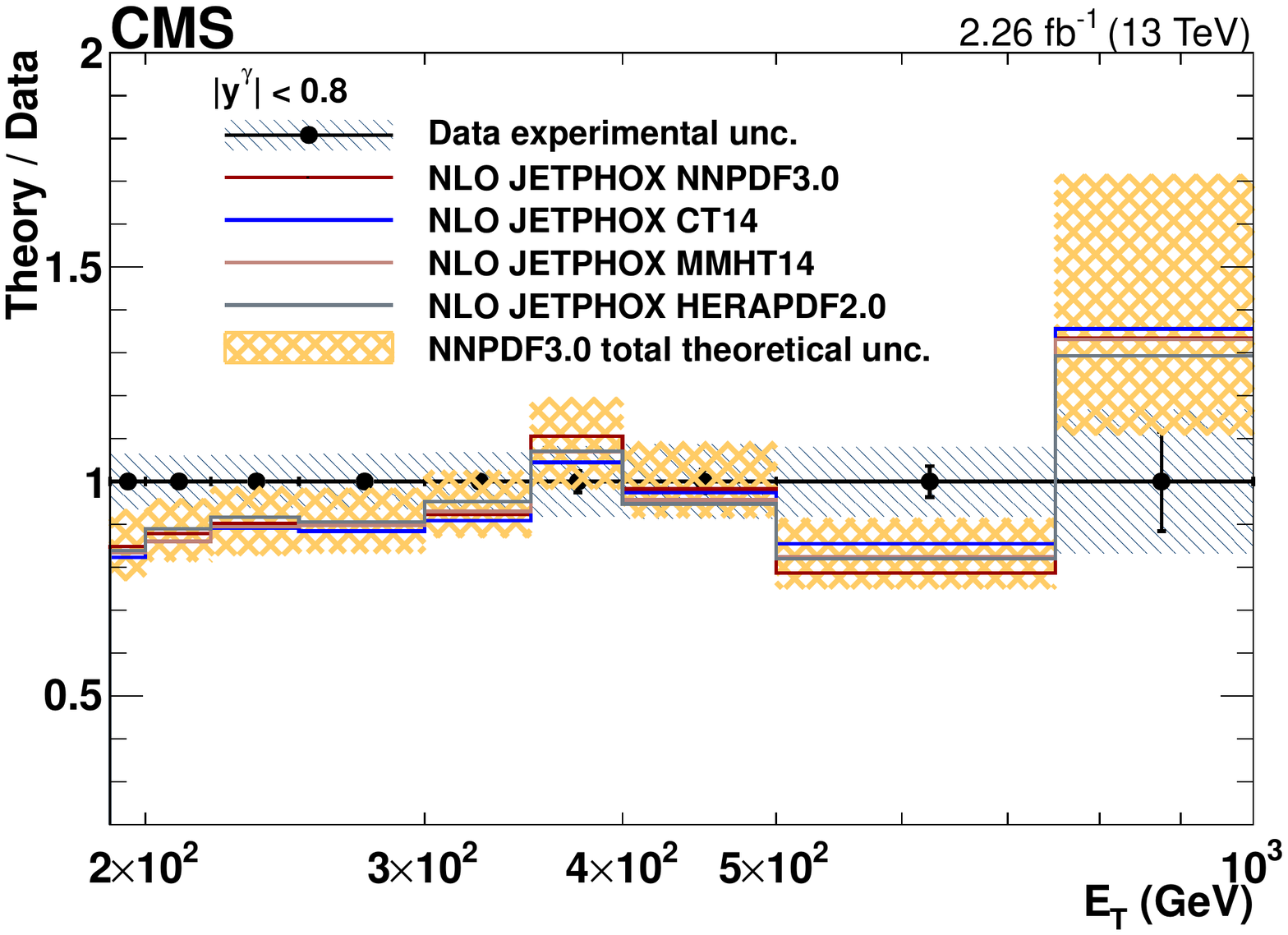}
\includegraphics[width=0.42\textwidth]{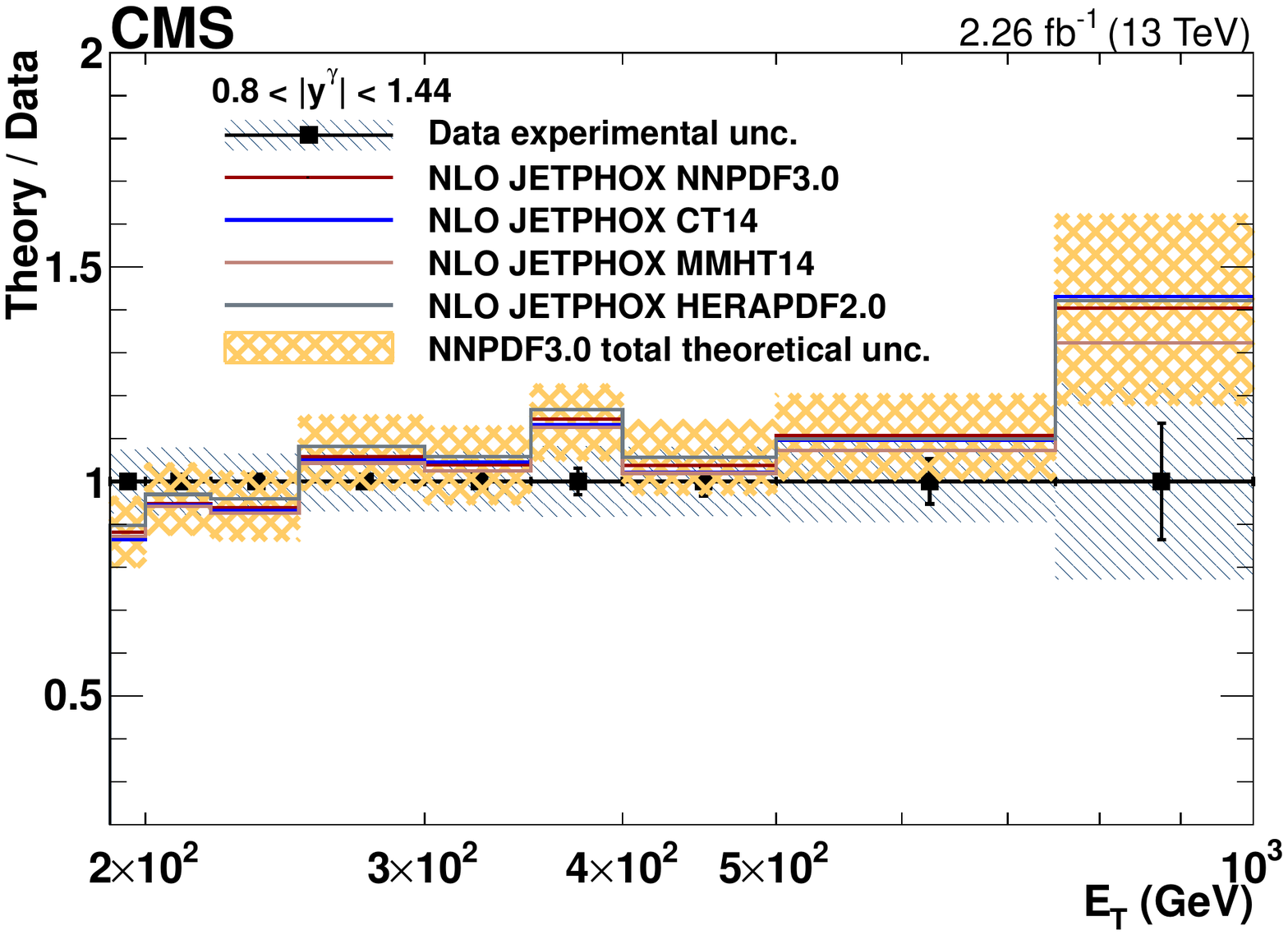}
\includegraphics[width=0.42\textwidth]{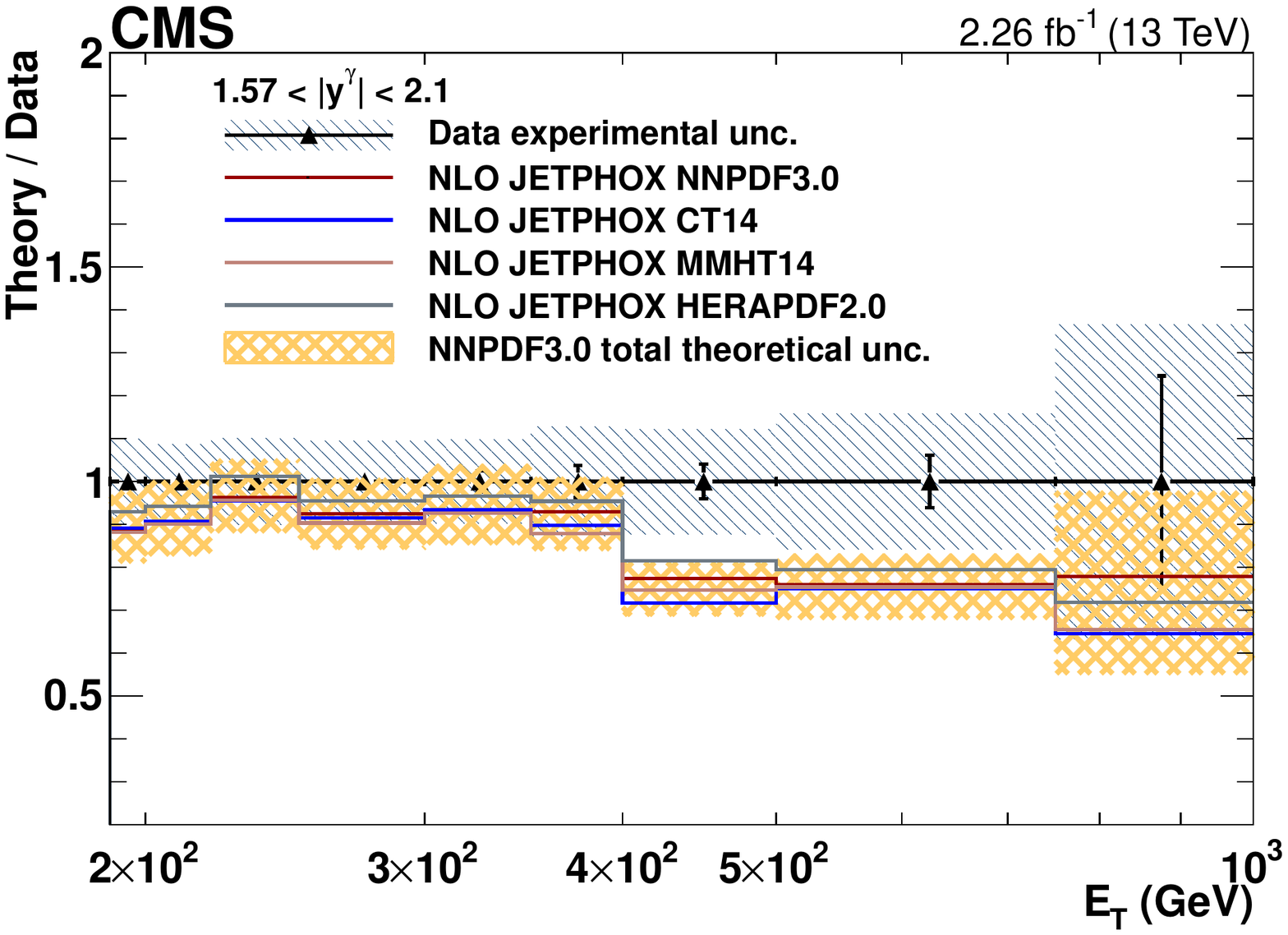}
\includegraphics[width=0.42\textwidth]{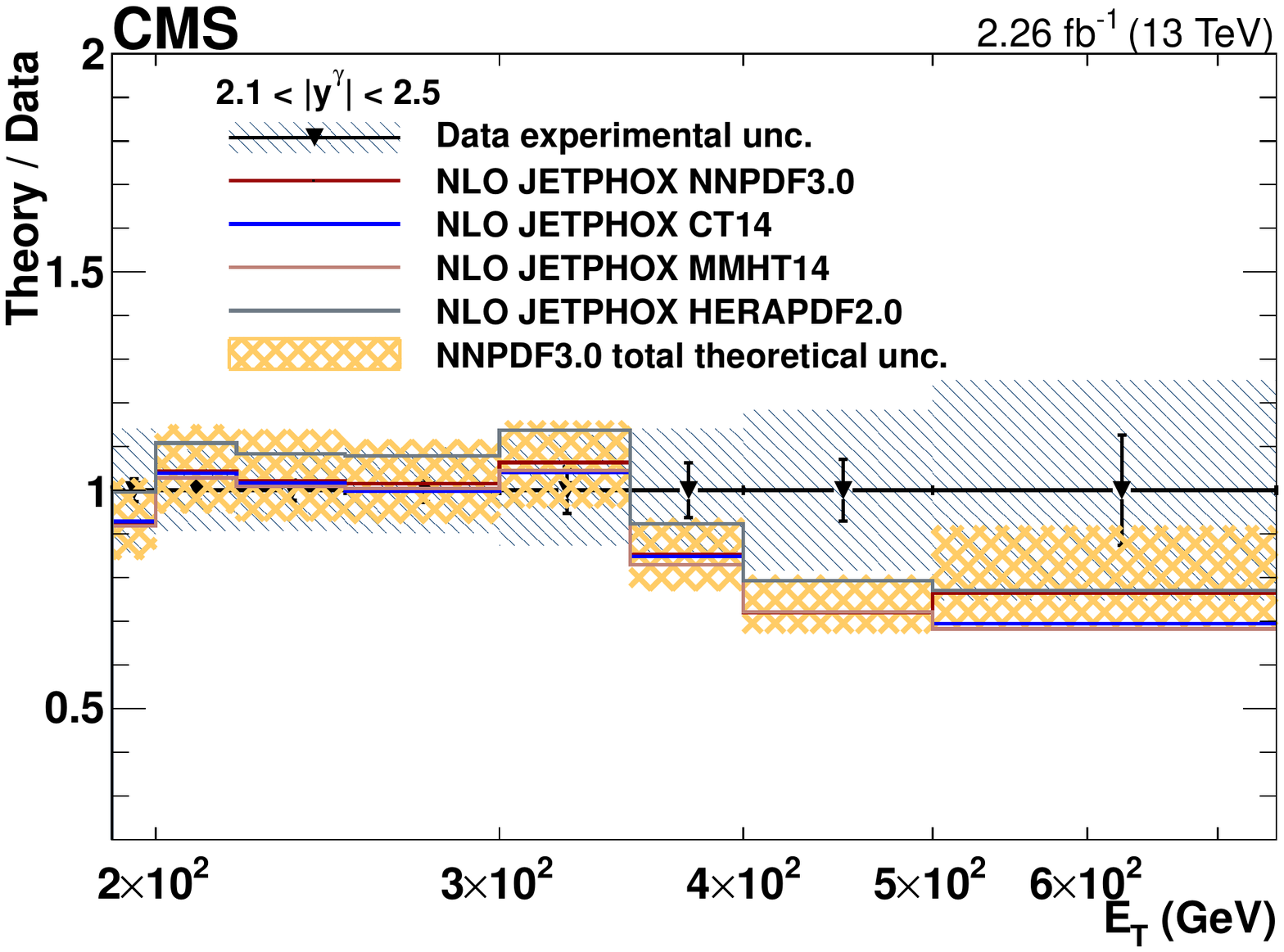}
\includegraphics[width=0.42\textwidth]{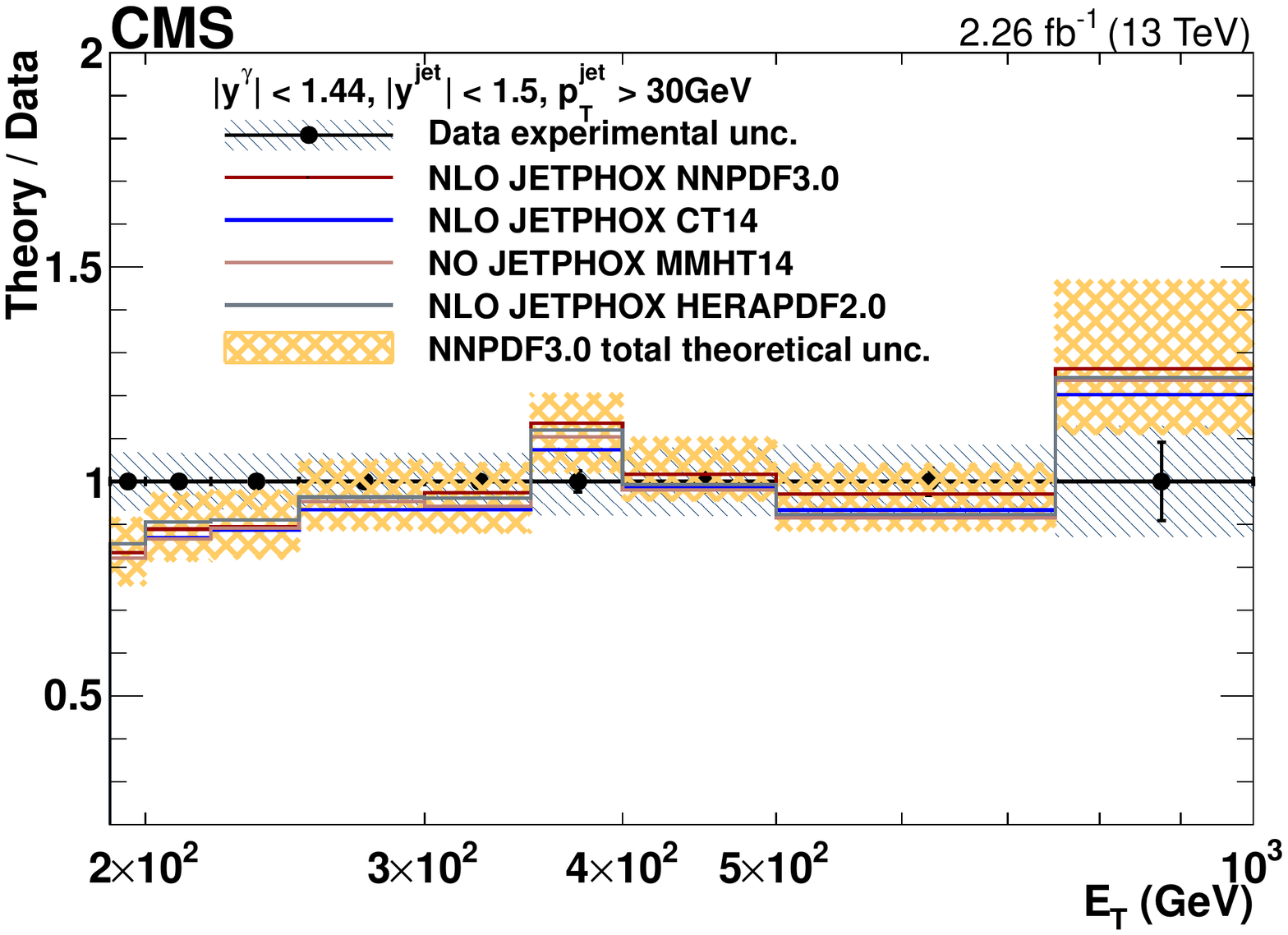}
\includegraphics[width=0.42\textwidth]{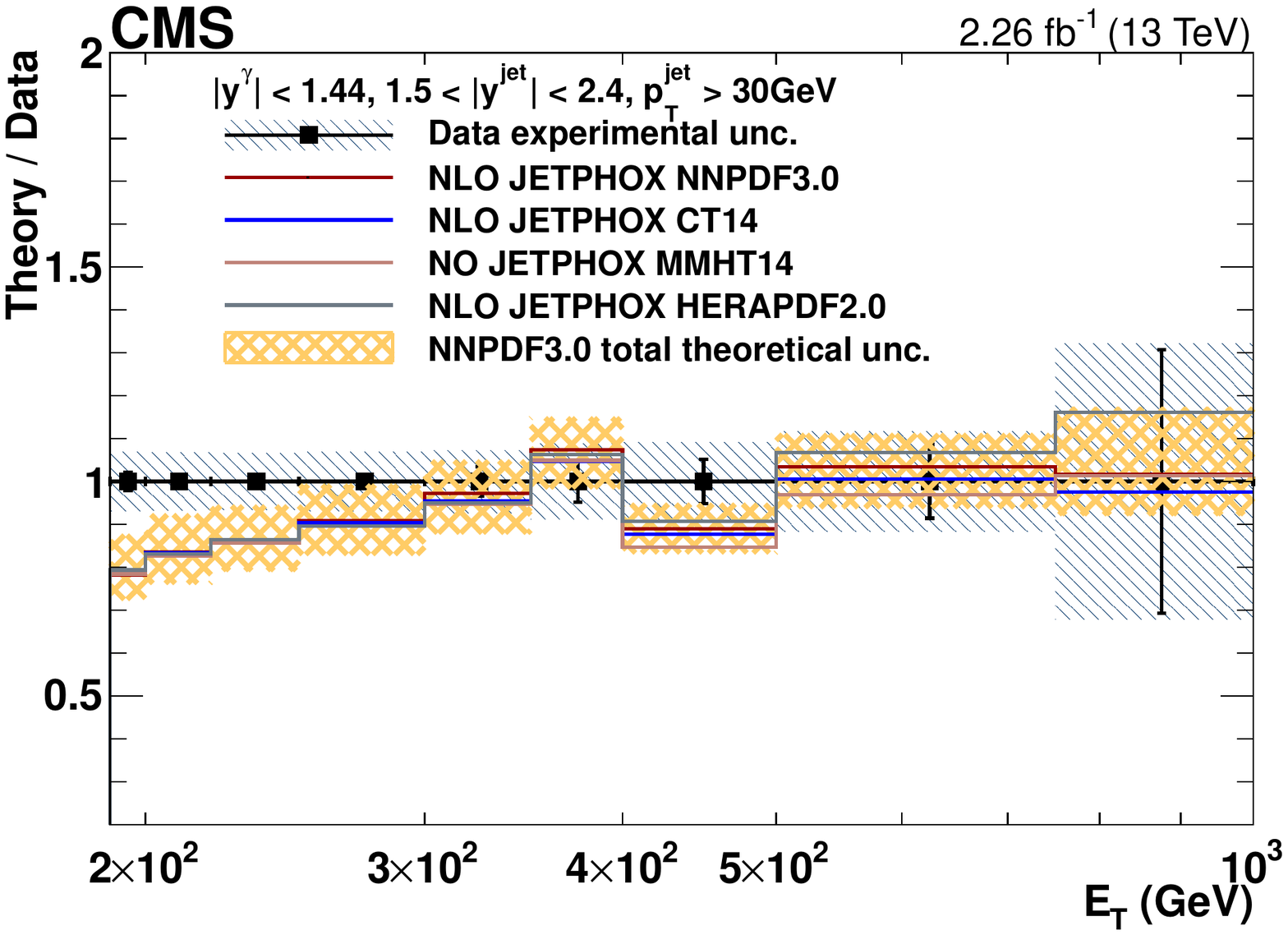}
\includegraphics[width=0.42\textwidth]{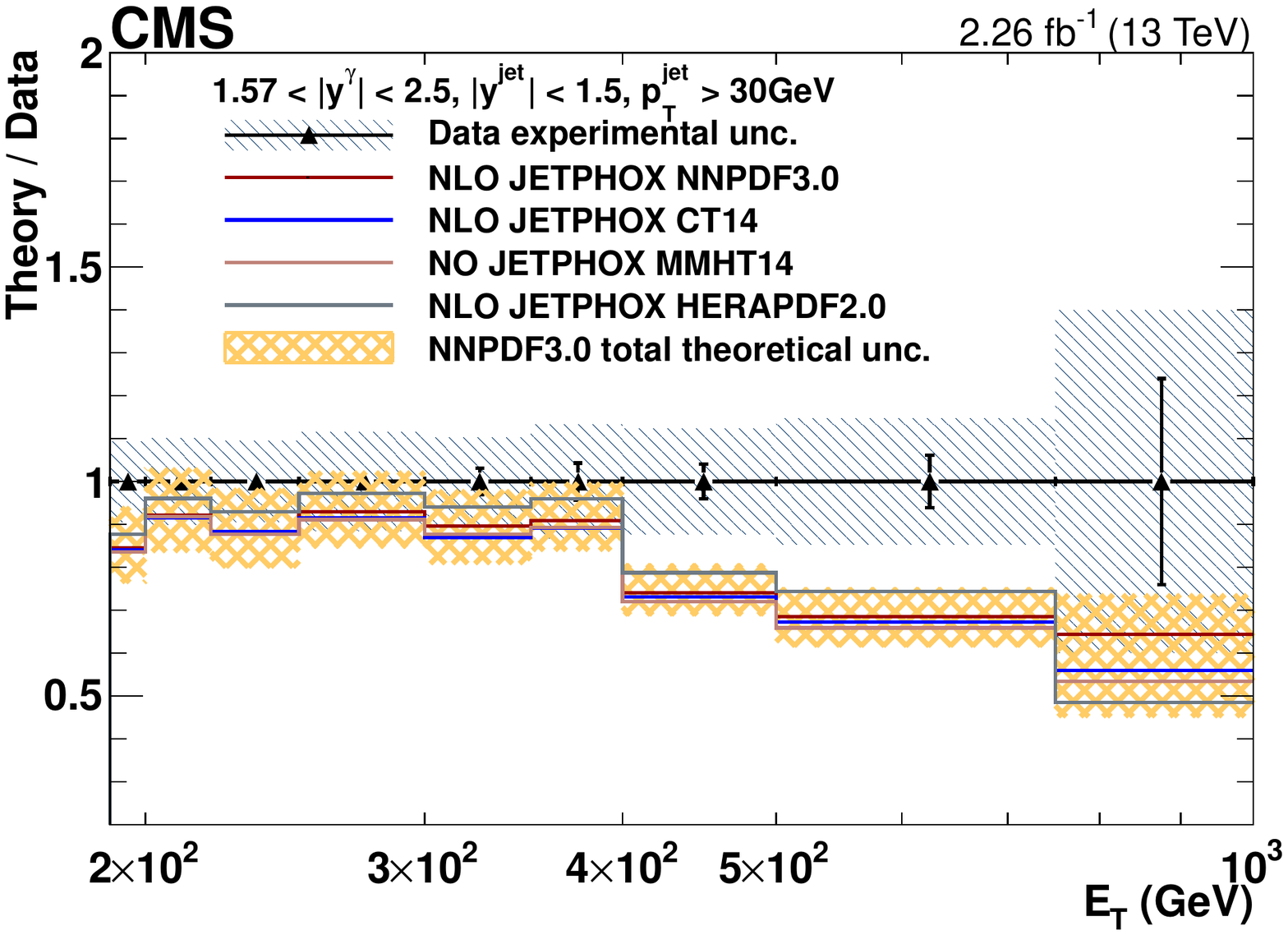}
\includegraphics[width=0.42\textwidth]{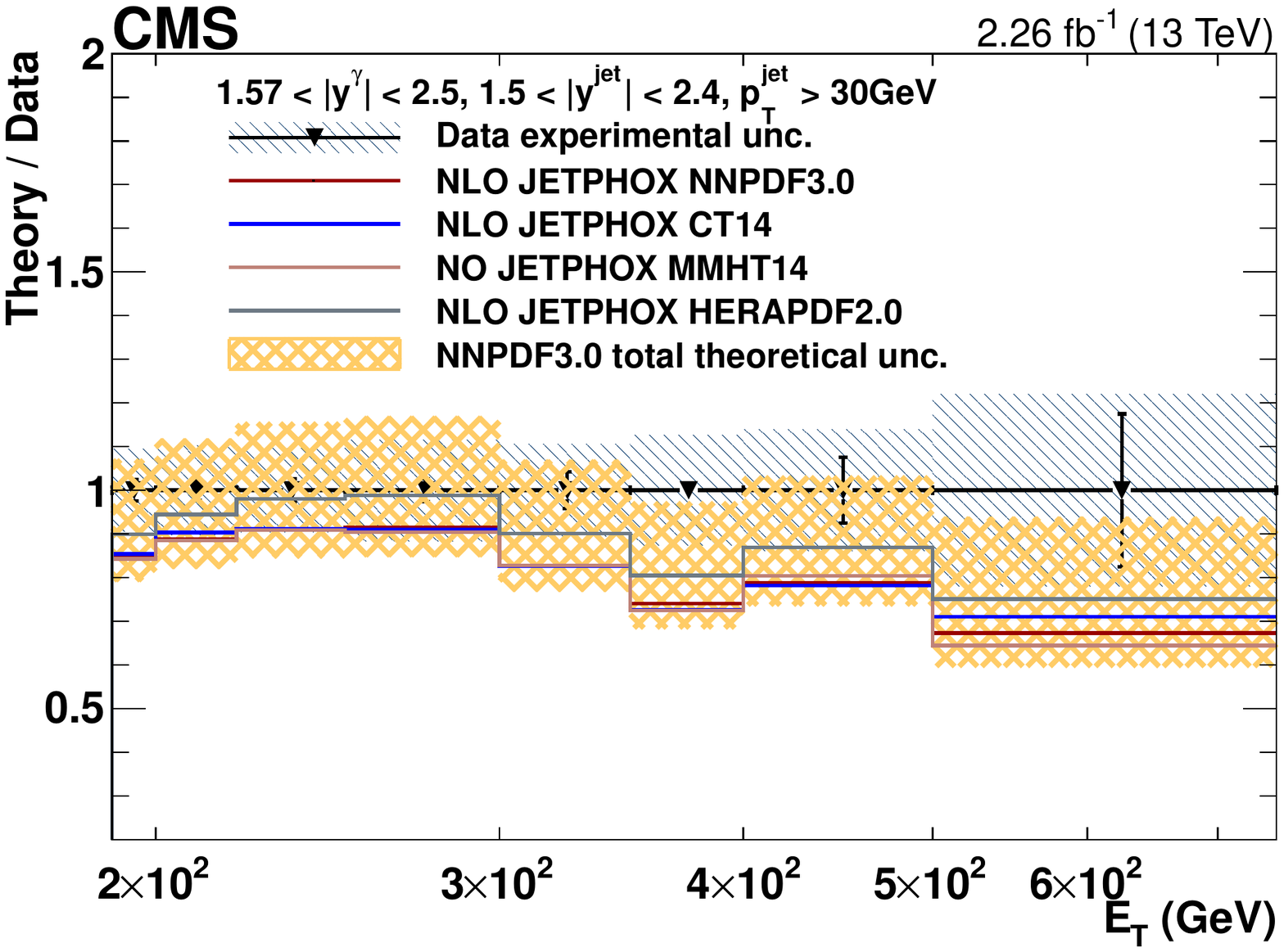}
\caption{Ratios of \JETPHOX~NLO predictions to data for various PDF sets as a function of photon \et for inclusive isolated-photons (top four panels)
and photon+jet (four bottom panels).
Data are shown as points, the error bars represent statistical uncertainties, while the hatched area represents the total experimental uncertainties.
The theoretical uncertainty in the NNPDF3.0 prediction is shown as a shaded area.}
\label{fig:phojet_XS_pdfs}
\end{figure*}

The ratio of the theoretical predictions to data based on \JETPHOX~at NLO with different PDF sets, including MMHT14~\cite{Harland-Lang:2014zoa}, CT14~\cite{Dulat:2015mca}, and HERAPDF2.0~\cite{Abramowicz:2015mha} together with NNPDF3.0, are shown in Fig.~\ref{fig:phojet_XS_pdfs}. The differences between \JETPHOX~predictions using different PDF sets are small, within the theoretical uncertainties estimated with NNPDF3.0.

\section{Summary}
The differential cross sections for inclusive isolated-photon and photon+jet production in proton-proton collisions at a center-of-mass energy of 13\TeV are measured with a data sample collected by the CMS experiment corresponding to an integrated luminosity of 2.26\fbinv.
The measurements of inclusive isolated-photon production cross sections are presented as functions of photon transverse energy and rapidity with $\et^{\gamma} > 190\GeV$ and $|y^{\gamma}|<2.5$.
The photon+jet production cross sections are presented as functions of photon transverse energy, and photon and jet rapidities, with requirement of an isolated photon and jet where $\pt^{\text{jet}}>30\GeV$ and $|y^{\text{jet}}|<2.4$.

The measurements are compared with theoretical predictions produced using the \JETPHOX~next-to-leading order calculations
using different parton distribution functions.
The theoretical predictions agree with the experimental measurements within the statistical and systematic uncertainties.
For low to middle range in photon $\et$, where the experimental uncertainties are smaller or comparable to theoretical uncertainties, these measurements provide the potential to further constrain the proton PDFs.  The agreement between data and theory, and the new next-to-next-to-leading-order (NNLO) calculations~\cite{Campbell:2016lzl} motivate the use of additional measurements to better estimate the gluon and other PDFs.

\begin{table*}[thp]
\topcaption{Measured and predicted differential cross section for isolated-photon production,
along with the statistical and systematical uncertainties in the various \et and $y$ bins.
Predictions use  \JETPHOX~at NLO with the NNPDF3.0 PDF set.
The ratio of the \JETPHOX~predictions to data are listed in the last column,
with the total uncertainty estimated assuming uncorrelated experimental and theoretical uncertainties.}
\label{tab:summary_xs}
\centering
\cmsTable{
\begin{tabular}{lccc}
\hline
\et(\GeVns{}) & Measured cross section within the bin (pb) & \specialcell{\JETPHOX\\NNPDF3.0 (pb)} & \JETPHOX/Data \\ \hline
\multicolumn{4}{c}{$\abs{y^{\gamma}}<0.8$} \\
190--200  &      $(3.64\pm0.04\stat\pm0.23\syst) \times 10^{-1}$ & ($3.1\pm0.3) \times 10^{-1}$  &  $0.85 \pm 0.10$  \\
200--220  &      $(2.49\pm0.02\stat\pm0.15\syst) \times 10^{-1}$ & ($2.2\pm0.2) \times 10^{-1}$  &  $0.88 \pm 0.09$  \\
220--250  &      $(1.46\pm0.01\stat\pm0.09\syst) \times 10^{-1}$ & ($1.3\pm0.1) \times 10^{-1}$  &  $0.90 \pm 0.10$  \\
250--300  &      $(7.09\pm0.08\stat\pm0.45\syst) \times 10^{-2}$ & ($6.4\pm0.5) \times 10^{-2}$  &  $0.91 \pm 0.10$  \\
300--350  &      $(2.91\pm0.05\stat\pm0.19\syst) \times 10^{-2}$ & ($2.7\pm0.3) \times 10^{-2}$  &  $0.92 \pm 0.12$  \\
350--400  &      $(1.24\pm0.03\stat\pm0.10\syst) \times 10^{-2}$ & ($1.4\pm0.2) \times 10^{-2}$  &  $1.11 \pm 0.15$  \\
400--500  &      $(5.1\pm0.1\stat\pm0.4\syst)    \times 10^{-3}$ & ($5.0\pm0.6) \times 10^{-3}$  &  $0.98 \pm 0.14$  \\
500--750  &      $(1.11\pm0.04\stat\pm0.08\syst) \times 10^{-3}$ & ($9.0\pm1.0) \times 10^{-4}$  &  $0.79 \pm 0.14$  \\
750--1000 &      $(1.0\pm0.1\stat\pm0.1\syst)    \times 10^{-4}$ & ($1.4\pm0.4) \times 10^{-4}$  &  $1.33 \pm 0.44$  \\ [\cmsTabSkip]
\multicolumn{4}{c}{$0.8<\abs{y^{\gamma}}<1.44$} \\
190--200  &      $(3.44\pm0.04\stat\pm0.25\syst) \times 10^{-1}$ & ($3.0\pm0.3) \times 10^{-1}$  &  $0.88 \pm 0.10$  \\
200--220  &      $(2.26\pm0.03\stat\pm0.18\syst) \times 10^{-1}$ & ($2.1\pm0.2) \times 10^{-1}$  &  $0.95 \pm 0.12$  \\
220--250  &      $(1.37\pm0.02\stat\pm0.09\syst) \times 10^{-1}$ & ($1.3\pm0.1) \times 10^{-1}$  &  $0.94 \pm 0.10$  \\
250--300  &      $(5.87\pm0.08\stat\pm0.40\syst) \times 10^{-2}$ & ($6.2\pm0.6) \times 10^{-2}$  &  $1.06 \pm 0.12$  \\
300--350  &      $(2.60\pm0.05\stat\pm0.17\syst) \times 10^{-2}$ & ($2.7\pm0.2) \times 10^{-2}$  &  $1.04 \pm 0.12$  \\
350--400  &      $(1.15\pm0.04\stat\pm0.09\syst) \times 10^{-2}$ & ($1.3\pm0.1) \times 10^{-2}$  &  $1.15 \pm 0.13$  \\
400--500  &      $(4.6\pm0.2\stat\pm0.3\syst)    \times 10^{-3}$ & ($4.7\pm0.5) \times 10^{-3}$  &  $1.04 \pm 0.13$  \\
500--750  &      $(7.4\pm0.4\stat\pm0.6\syst)    \times 10^{-4}$ & ($8.2\pm0.8) \times 10^{-4}$  &  $1.11 \pm 0.15$  \\
750--1000 &      $(8.0\pm1.0\stat\pm1.0\syst)    \times 10^{-5}$ & ($1.1\pm0.2) \times 10^{-4}$  &  $1.40 \pm 0.39$  \\ [\cmsTabSkip]
\multicolumn{4}{c}{$1.57<\abs{y^{\gamma}}<2.1$} \\
190--200  &      $(3.16\pm0.05\stat\pm0.31\syst) \times 10^{-1}$ & ($2.8\pm0.3) \times 10^{-1}$  &  $0.88 \pm 0.13$  \\
200--220  &      $(2.19\pm0.03\stat\pm0.19\syst) \times 10^{-1}$ & ($2.0\pm0.2) \times 10^{-1}$  &  $0.91 \pm 0.12$  \\
220--250  &      $(1.19\pm0.02\stat\pm0.12\syst) \times 10^{-1}$ & ($1.1\pm0.1) \times 10^{-1}$  &  $0.96 \pm 0.13$  \\
250--300  &      $(5.80\pm0.09\stat\pm0.54\syst) \times 10^{-2}$ & ($5.4\pm0.5) \times 10^{-2}$  &  $0.92 \pm 0.12$  \\
300--350  &      $(2.37\pm0.06\stat\pm0.22\syst) \times 10^{-2}$ & ($2.2\pm0.3) \times 10^{-2}$  &  $0.93 \pm 0.14$  \\
350--400  &      $(1.02\pm0.04\stat\pm0.12\syst) \times 10^{-2}$ & ($9.5\pm0.9) \times 10^{-3}$  &  $0.93 \pm 0.15$  \\
400--500  &      $(4.0\pm0.2\stat\pm0.5\syst)    \times 10^{-3}$ & ($3.1\pm0.3) \times 10^{-3}$  &  $0.77 \pm 0.13$  \\
500--750  &      $(6.1\pm0.4\stat\pm0.9\syst)    \times 10^{-4}$ & ($4.6\pm0.5) \times 10^{-4}$  &  $0.76 \pm 0.14$  \\
750--1000 &      $(3.9\pm1.0\stat\pm1.1\syst)    \times 10^{-5}$ & ($3.0\pm0.9) \times 10^{-5}$  &  $0.78 \pm 0.37$  \\ [\cmsTabSkip]
\multicolumn{4}{c}{$2.1<\abs{y^{\gamma}}<2.5$} \\
190--200  &      $(2.52\pm0.07\stat\pm0.35\syst) \times 10^{-1}$ & ($2.3\pm0.3) \times 10^{-1}$  &  $0.92 \pm 0.17$  \\
200--220  &      $(1.55\pm0.04\stat\pm0.14\syst) \times 10^{-1}$ & ($1.6\pm0.2) \times 10^{-1}$  &  $1.04 \pm 0.14$  \\
220--250  &      $(8.8\pm0.2\stat\pm0.8\syst)    \times 10^{-2}$ & ($9.0\pm1.0) \times 10^{-2}$  &  $1.02 \pm 0.15$  \\
250--300  &      $(3.7\pm0.1\stat\pm0.4\syst)    \times 10^{-2}$ & ($3.8\pm0.4) \times 10^{-2}$  &  $1.01 \pm 0.14$  \\
300--350  &      $(1.32\pm0.07\stat\pm0.15\syst) \times 10^{-2}$ & ($1.4\pm0.1) \times 10^{-2}$  &  $1.06 \pm 0.17$  \\
350--400  &      $(5.9\pm0.4\stat\pm0.7\syst)    \times 10^{-3}$ & ($5.0\pm0.5) \times 10^{-3}$  &  $0.85 \pm 0.14$  \\
400--500  &      $(1.7\pm0.1\stat\pm0.3\syst)    \times 10^{-3}$ & ($1.2\pm0.1) \times 10^{-3}$  &  $0.72 \pm 0.16$  \\
500--750  &      $(1.8\pm0.2\stat\pm0.4\syst)    \times 10^{-4}$ & ($1.4\pm0.3) \times 10^{-4}$  &  $0.77 \pm 0.25$  \\
\hline
\end{tabular}
}
\end{table*}

\begin{table*}[thp]
\topcaption{Measured and predicted differential cross section for photon+jet production, along with statistical and
systematical uncertainties in the various \et and $y$ bins.
Predictions are based on \JETPHOX~at NLO with the NNPDF3.0 PDF set.
The ratio of the \JETPHOX~predictions to the data are listed in the last column,
with the total uncertainty estimated assuming uncorrelated experimental and theoretical uncertainties.}
\label{tab:summary_xs_phojet}
\centering
\cmsTable{
\begin{tabular}{lccc}
\hline
\et(\GeVns{}) & Measured cross section within the bin (pb) & \specialcell{\JETPHOX\\NNPDF3.0 (pb)} &  \JETPHOX/Data\\ \hline
\multicolumn{4}{c}{$\abs{y^{\gamma}}<1.44$, $\abs{y^{\text{jet}}}<1.5$, and $\pt^{\text{jet}}>30\GeV$} \\
190--200  &      $(9.2\pm0.1\stat\pm0.6\syst)    \times 10^{-2}$ & ($7.7\pm0.7) \times 10^{-2}$  &  $0.83 \pm 0.10$  \\
200--220  &      $(6.26\pm0.06\stat\pm0.41\syst) \times 10^{-2}$ & ($5.6\pm0.5) \times 10^{-2}$  &  $0.89 \pm 0.10$  \\
220--250  &      $(3.72\pm0.04\stat\pm0.23\syst) \times 10^{-2}$ & ($3.3\pm0.3) \times 10^{-2}$  &  $0.89 \pm 0.10$  \\
250--300  &      $(1.72\pm0.02\stat\pm0.11\syst) \times 10^{-2}$ & ($1.6\pm0.2) \times 10^{-2}$  &  $0.95 \pm 0.12$  \\
300--350  &      $(7.50\pm0.1\stat\pm0.5\syst)   \times 10^{-3}$ & ($7.3\pm0.7) \times 10^{-3}$  &  $0.97 \pm 0.11$  \\
350--400  &      $(3.34\pm0.08\stat\pm0.25\syst) \times 10^{-3}$ & ($3.8\pm0.4) \times 10^{-3}$  &  $1.14 \pm 0.15$  \\
400--500  &      $(1.37\pm0.03\stat\pm0.10\syst) \times 10^{-3}$ & ($1.4\pm0.1) \times 10^{-3}$  &  $1.02 \pm 0.12$  \\
500--750  &      $(2.82\pm0.09\stat\pm0.22\syst) \times 10^{-4}$ & ($2.7\pm0.2) \times 10^{-4}$  &  $0.97 \pm 0.12$  \\
750--1000 &      $(3.0\pm0.3\stat\pm0.3\syst)    \times 10^{-5}$ & ($3.8\pm0.6) \times 10^{-5}$  &  $1.26 \pm 0.26$  \\ [\cmsTabSkip]
\multicolumn{4}{c}{$\abs{y^{\gamma}}<1.44$, $1.5<\abs{y^{\text{jet}}}<2.4$, and $\pt^{\text{jet}}>30\GeV$} \\
190--200  &      $(4.08\pm0.09\stat\pm0.27\syst) \times 10^{-2}$ & ($3.2\pm0.4) \times 10^{-2}$  &  $0.78 \pm 0.11$  \\
200--220  &      $(2.73\pm0.05\stat\pm0.18\syst) \times 10^{-2}$ & ($2.3\pm0.2) \times 10^{-2}$  &  $0.84 \pm 0.10$  \\
220--250  &      $(1.54\pm0.03\stat\pm0.10\syst) \times 10^{-2}$ & ($1.3\pm0.1) \times 10^{-2}$  &  $0.86 \pm 0.10$  \\
250--300  &      $(6.9\pm0.1\stat\pm0.5\syst)    \times 10^{-3}$ & ($6.3\pm0.6) \times 10^{-3}$  &  $0.91 \pm 0.10$  \\
300--350  &      $(2.73\pm0.09\stat\pm0.18\syst) \times 10^{-3}$ & ($2.7\pm0.3) \times 10^{-3}$  &  $0.97 \pm 0.12$  \\
350--400  &      $(1.12\pm0.05\stat\pm0.08\syst) \times 10^{-3}$ & ($1.2\pm0.1) \times 10^{-3}$  &  $1.07 \pm 0.13$  \\
400--500  &      $(4.4\pm0.2\stat\pm0.3\syst)    \times 10^{-4}$ & ($3.9\pm0.3) \times 10^{-4}$  &  $0.89 \pm 0.10$  \\
500--750  &      $(5.8\pm0.5\stat\pm0.5\syst)    \times 10^{-5}$ & ($6.0\pm0.6) \times 10^{-5}$  &  $1.03 \pm 0.15$  \\
750--1000 &      $(4.3\pm1.3\stat\pm0.4\syst)    \times 10^{-6}$ & ($4.4\pm0.7) \times 10^{-6}$  &  $1.02 \pm 0.36$  \\ [\cmsTabSkip]
\multicolumn{4}{c}{$1.57<\abs{y^{\gamma}}<2.5$, $\abs{y^{\text{jet}}}<1.5$, and $\pt^{\text{jet}}>30\GeV$} \\
190--200  &      $(6.0\pm0.1\stat\pm0.6\syst)    \times 10^{-2}$ & ($5.1\pm0.6) \times 10^{-2}$  &  $0.85 \pm 0.12$  \\
200--220  &      $(3.92\pm0.08\stat\pm0.39\syst) \times 10^{-2}$ & ($3.6\pm0.4) \times 10^{-2}$  &  $0.92 \pm 0.14$  \\
220--250  &      $(2.42\pm0.04\stat\pm0.23\syst) \times 10^{-2}$ & ($2.1\pm0.2) \times 10^{-2}$  &  $0.88 \pm 0.13$  \\
250--300  &      $(1.08\pm0.02\stat\pm0.12\syst) \times 10^{-2}$ & ($1.0\pm0.1) \times 10^{-2}$  &  $0.93 \pm 0.14$  \\
300--350  &      $(4.7\pm0.1\stat\pm0.5\syst)    \times 10^{-3}$ & ($4.2\pm0.4) \times 10^{-3}$  &  $0.90 \pm 0.13$  \\
350--400  &      $(2.03\pm0.09\stat\pm0.25\syst) \times 10^{-3}$ & ($1.8\pm0.2) \times 10^{-3}$  &  $0.91 \pm 0.15$  \\
400--500  &      $(8.1\pm0.3\stat\pm0.9\syst)    \times 10^{-4}$ & ($6.0\pm0.5) \times 10^{-4}$  &  $0.74 \pm 0.11$  \\
500--750  &      $(1.24\pm0.08\stat\pm0.17\syst) \times 10^{-4}$ & ($8.5\pm0.9) \times 10^{-5}$  &  $0.69 \pm 0.12$  \\
750--1000 &      $(1.0\pm0.2\stat\pm0.3\syst)    \times 10^{-5}$ & ($6.0\pm2.0) \times 10^{-6}$  &  $0.64 \pm 0.32$  \\ [\cmsTabSkip]
\multicolumn{4}{c}{$1.57<\abs{y^{\gamma}}<2.5$, $1.5<\abs{y^{\text{jet}}}<2.4$, and $\pt^{\text{jet}}>30\GeV$} \\
190--200  &      $(5.0\pm0.1\stat\pm0.5\syst)    \times 10^{-2}$ & ($4.0\pm1.0) \times 10^{-2}$  &  $0.85 \pm 0.23$  \\
200--220  &      $(3.39\pm0.08\stat\pm0.34\syst) \times 10^{-2}$ & ($3.0\pm0.8) \times 10^{-2}$  &  $0.89 \pm 0.24$  \\
220--250  &      $(1.87\pm0.05\stat\pm0.17\syst) \times 10^{-2}$ & ($1.7\pm0.5) \times 10^{-2}$  &  $0.91 \pm 0.26$  \\
250--300  &      $(8.1\pm0.2\stat\pm0.9\syst)    \times 10^{-3}$ & ($7.0\pm2.0) \times 10^{-3}$  &  $0.92 \pm 0.27$  \\
300--350  &      $(3.4\pm0.1\stat\pm0.3\syst)    \times 10^{-3}$ & ($2.8\pm0.8) \times 10^{-3}$  &  $0.83 \pm 0.26$  \\
350--400  &      $(1.38\pm0.02\stat\pm0.17\syst) \times 10^{-3}$ & ($1.0\pm0.3) \times 10^{-3}$  &  $0.74 \pm 0.25$  \\
400--500  &      $(3.4\pm0.3\stat\pm0.4\syst)    \times 10^{-4}$ & ($2.7\pm0.8) \times 10^{-4}$  &  $0.79 \pm 0.27$  \\
500--750  &      $(4.1\pm0.7\stat\pm0.5\syst)    \times 10^{-5}$ & ($3.0\pm1.0) \times 10^{-5}$  &  $0.67 \pm 0.30$  \\
\hline
\end{tabular}
}
\end{table*}

\clearpage
\begin{acknowledgments}
We congratulate our colleagues in the CERN accelerator departments for the excellent performance of the LHC and thank the technical and administrative staffs at CERN and at other CMS institutes for their contributions to the success of the CMS effort. In addition, we gratefully acknowledge the computing centres and personnel of the Worldwide LHC Computing Grid for delivering so effectively the computing infrastructure essential to our analyses. Finally, we acknowledge the enduring support for the construction and operation of the LHC and the CMS detector provided by the following funding agencies: BMWFW and FWF (Austria); FNRS and FWO (Belgium); CNPq, CAPES, FAPERJ, and FAPESP (Brazil); MES (Bulgaria); CERN; CAS, MoST, and NSFC (China); COLCIENCIAS (Colombia); MSES and CSF (Croatia); RPF (Cyprus); SENESCYT (Ecuador); MoER, ERC IUT, and ERDF (Estonia); Academy of Finland, MEC, and HIP (Finland); CEA and CNRS/IN2P3 (France); BMBF, DFG, and HGF (Germany); GSRT (Greece); OTKA and NIH (Hungary); DAE and DST (India); IPM (Iran); SFI (Ireland); INFN (Italy); MSIP and NRF (Republic of Korea); LAS (Lithuania); MOE and UM (Malaysia); BUAP, CINVESTAV, CONACYT, LNS, SEP, and UASLP-FAI (Mexico); MBIE (New Zealand); PAEC (Pakistan); MSHE and NSC (Poland); FCT (Portugal); JINR (Dubna); MON, RosAtom, RAS, and RFBR (Russia); MESTD (Serbia); SEIDI and CPAN (Spain); Swiss Funding Agencies (Switzerland); MST (Taipei); ThEPCenter, IPST, STAR, and NSTDA (Thailand); TUBITAK and TAEK (Turkey); NASU and SFFR (Ukraine); STFC (United Kingdom); DOE and NSF (USA).

\hyphenation{Rachada-pisek} Individuals have received support from the Marie-Curie programme and the European Research Council and Horizon 2020 Grant, contract No. 675440 (European Union); the Leventis Foundation; the A. P. Sloan Foundation; the Alexander von Humboldt Foundation; the Belgian Federal Science Policy Office; the Fonds pour la Formation \`a la Recherche dans l'Industrie et dans l'Agriculture (FRIA-Belgium); the Agentschap voor Innovatie door Wetenschap en Technologie (IWT-Belgium); the F.R.S.-FNRS and FWO (Belgium) under the ``Excellence of Science - EOS" - be.h project n. 30820817; the Ministry of Education, Youth and Sports (MEYS) of the Czech Republic; the Lend\"ulet (``Momentum") Programme and the J\'anos Bolyai Research Scholarship of the Hungarian Academy of Sciences, the New National Excellence Program \'UNKP, the NKFIA research grants 123842, 123959, 124845, 124850 and 125105 (Hungary); the Council of Science and Industrial Research, India; the HOMING PLUS programme of the Foundation for Polish Science, cofinanced from European Union, Regional Development Fund, the Mobility Plus programme of the Ministry of Science and Higher Education, the National Science Center (Poland), contracts Harmonia 2014/14/M/ST2/00428, Opus 2014/13/B/ST2/02543, 2014/15/B/ST2/03998, and 2015/19/B/ST2/02861, Sonata-bis 2012/07/E/ST2/01406; the National Priorities Research Program by Qatar National Research Fund; the Programa Estatal de Fomento de la Investigaci{\'o}n Cient{\'i}fica y T{\'e}cnica de Excelencia Mar\'{\i}a de Maeztu, grant MDM-2015-0509 and the Programa Severo Ochoa del Principado de Asturias; the Thalis and Aristeia programmes cofinanced by EU-ESF and the Greek NSRF; the Rachadapisek Sompot Fund for Postdoctoral Fellowship, Chulalongkorn University and the Chulalongkorn Academic into Its 2nd Century Project Advancement Project (Thailand); the Welch Foundation, contract C-1845; and the Weston Havens Foundation (USA).
\end{acknowledgments}

\bibliography{auto_generated}

\cleardoublepage \appendix\section{The CMS Collaboration \label{app:collab}}\begin{sloppypar}\hyphenpenalty=5000\widowpenalty=500\clubpenalty=5000\vskip\cmsinstskip
\textbf{Yerevan Physics Institute, Yerevan, Armenia}\\*[0pt]
A.M.~Sirunyan, A.~Tumasyan
\vskip\cmsinstskip
\textbf{Institut f\"{u}r Hochenergiephysik, Wien, Austria}\\*[0pt]
W.~Adam, F.~Ambrogi, E.~Asilar, T.~Bergauer, J.~Brandstetter, E.~Brondolin, M.~Dragicevic, J.~Er\"{o}, A.~Escalante~Del~Valle, M.~Flechl, R.~Fr\"{u}hwirth\cmsAuthorMark{1}, V.M.~Ghete, J.~Hrubec, M.~Jeitler\cmsAuthorMark{1}, N.~Krammer, I.~Kr\"{a}tschmer, D.~Liko, T.~Madlener, I.~Mikulec, N.~Rad, H.~Rohringer, J.~Schieck\cmsAuthorMark{1}, R.~Sch\"{o}fbeck, M.~Spanring, D.~Spitzbart, A.~Taurok, W.~Waltenberger, J.~Wittmann, C.-E.~Wulz\cmsAuthorMark{1}, M.~Zarucki
% ...
\vskip\cmsinstskip
% ...
1:  Also at Vienna University of Technology, Vienna, Austria\\
% ...
\end{sloppypar}
\end{document}